\documentclass[a4paper,11pt]{article}
\pdfoutput=1

\usepackage{jcappub}
\usepackage[T1]{fontenc}
\renewcommand{\Re}{\operatorname{Re}}
\renewcommand{\Im}{\operatorname{Im}}

\title{\boldmath Inflationary $\alpha$-Attractor Models with Singular Derivative of Potential}

\author[a]{Kei-ichiro Kubota,}
\author[a]{Hiroki Matsui,}
\author[b]{Takahiro Terada}

\affiliation[a]{Center for Gravitational Physics and Quantum Information, Yukawa Institute for Theoretical Physics, Kyoto University, Kitashirakawa Oiwakecho, Sakyo-ku, Kyoto 606-8502, Japan}
\affiliation[b]{Center for Theoretical Physics of the Universe, Institute for Basic Science (IBS), \\ 55, Expo-ro, Yuseong-gu, Daejeon, 34126, Korea}

\emailAdd{keiichiro.kubota@yukawa.kyoto-u.ac.jp}
\emailAdd{hiroki.matsui@yukawa.kyoto-u.ac.jp}
\emailAdd{takahiro.terada.hepc@gmail.com}

\keywords{inflation, primordial black hole, gravitational wave}

\abstract{A generalization of inflationary $\alpha$-attractor models (``polynomial $\alpha$-attractor'') was recently proposed by Kallosh and Linde, in which the potential involves logarithmic functions of the inflaton so that the derivative of the potential but not potential itself has a singularity.
We find that the models can lead to viable inflationary observables even without the pole in the kinetic term.
Also, the generalization with a pole order other than two does not significantly change the functional form of the potential.
This allows a systematic analysis of the predictions of this class of models. 
Our models predict larger spectral index $n_s$ and tensor-to-scalar ratio $r$ than in the polynomial $\alpha$-attractor: typically, $n_s$ around 0.97--0.98 and $r$ observable by LiteBIRD.
Taking advantage of the relatively large $n_s$, we discuss the modification of the potential to produce primordial black holes as the whole dark matter and gravitational waves induced by curvature perturbations detectable by LISA and BBO/DECIGO, while keeping $n_s$ in agreement with the Planck/BICEP/Keck data.
}

\begin{document}

\vspace{ 0cm}
\hfill {\small YITP-23-15}\\
\vspace{ 0cm}
\hfill {\small CTPU-PTC-23-03}\\

\maketitle
\flushbottom

%%%%%%%%%%%%%%%%%%%%%%%%%%%%%%%%%%%%%%%%%%%%
\section{Introduction \label{sec:intro}}
%%%%%%%%%%%%%%%%%%%%%%%%%%%%%%%%%%%%%%%%%%%%

Recent analyses of the cosmic microwave background (CMB) data put constraints on inflationary observables with unprecedented precision.   For example, the upper bound on the tensor-to-scalar ratio by BICEP/Keck collaboration has become $r < 0.036$~\cite{BICEP:2021xfz} at 95\% confidence level or $r < 0.032$~\cite{Tristram:2021tvh} when combined with the latest Planck data~\cite{Planck:2020olo}.  In view of the potential sensitivity of $\delta r  < 0.001$~\cite{LiteBIRD:2020khw} for the next-generation observation by LiteBIRD, it is important to thoroughly explore the landscape of inflation models that predict $r \sim \mathcal{O}(10^{-3} \text{--} 10^{-2})$.  

In addition to the $B$-mode signals from the tensor-to-scalar ratio, it is also important to precisely measure the scalar spectral index $n_s$ to narrow down various possibilities of the underlying inflation models.  A class of models including Higgs inflation~\cite{Bezrukov:2007ep}, Starobinsky model~\cite{Starobinsky:1980te}, and $\alpha$-attractor models of inflation~\cite{Ellis:2013nxa, Ferrara:2013rsa, Kallosh:2013yoa, Kallosh:2014rga, Carrasco:2015pla, Roest:2015qya, Linde:2015uga, Scalisi:2015qga} predicts $1 - n_s =  2 /N_e $ at the leading order of $1/N_e $ expansion where $N_e $ is the e-folding number, which is successful in fitting the CMB data.  The characteristics of this class of models can be understood in a unified manner by focusing on the second-order pole in the inflaton kinetic term in the Einstein frame~\cite{Galante:2014ifa}. Such a pole can be further traced back, e.g., to the non-minimal coupling between the inflaton and the scalar curvature term in the Jordan frame or to an analogous term in the superconformal setting.   

Typically, small-field models (such as hilltop inflation models~\cite{Linde:1981mu, Albrecht:1982wi, Boubekeur:2005zm}) tend to predict smaller values of $n_s$, while other large field models (in particular, inverse-hilltop types) tend to predict larger values of $n_s$.  In fact, these inflationary models can be classified by the large $N_e$ behaviors of their predictions on inflationary observables (the universality classes of inflation~\cite{Mukhanov:2013tua, Roest:2013fha, Garcia-Bellido:2014gna, Binetruy:2014zya}).  Meanwhile, generalization of $\alpha$-attractor models of inflation to models with a higher- or lower-order pole in the kinetic term leads to a realization of some universality classes of inflation~\cite{Galante:2014ifa, Broy:2015qna}.  

In supergravity setting, it is not automatically guaranteed to have a singularity only in the kinetic term but not in the scalar potential.  In this sense, it is natural to consider the effects of a potentially existing singularity in the inflaton potential.  The presence of singularities both in the kinetic and potential terms leads to a richer variety of universality classes of inflation~\cite{Terada:2016nqg}.  

Recently, another generalization of the $\alpha$-attractor models was proposed~\cite{Kallosh:2022feu}, in which the scalar potential does not have a singularity, but its first derivative has a singularity whose position in the field space coincides with the position of the second-order pole in the kinetic term.  It can be realized, e.g.,\footnote{
Another possibility to realize a singularity in the first derivative (but not for the potential itself) is to use a fractional power potential.  However, this does not lead to qualitatively different results compared to simple monomial chaotic inflation models~\cite{Linde:1983gd}, so we do not focus on such possibilities.
} with the potential $V(\rho) = V_0\frac{(\ln \rho)^{2n}}{(\ln \rho)^{2n} + 1}$~\cite{Kallosh:2022feu}
where $n$ is a positive integer.  By proceeding with canonical normalization, this potential 
at large $\rho\sim \exp(-\phi)$ morphs into the inverse-hilltop potential, $V(\phi) \sim V_0( 1 - \frac{\mu^{2n}}{\phi^{2n}} + \cdots)$, rather than the standard exponential form, and that is why the model was called the ``polynomial $\alpha$-attractor''.
The polynomial $\alpha$-attractor opens the possibility of realizing larger values of $n_s$ in the $\alpha$-attractor models and covers the right side of the central value ($n_s\gtrsim 0.97$) of the Planck/BICEP/Keck data~\cite{Kallosh:2022feu,Bhattacharya:2022akq}.
Such an increase of $n_s$ has advantages\footnote{The observation of Atacama Cosmology Telescope (ACT) found $n_s  \simeq 1$ (the Harrison-Zeldovich spectrum)~\cite{ACT:2020gnv}. Some authors proposed that such a drastic change of the spectral index is necessary to solve the Hubble tension~\cite{Braglia:2020bym,Jiang:2022uyg,Cruz:2022oqk}.} in scenarios such as two-step inflation~\cite{Silk:1986vc} and primordial black hole (PBH) formation~\cite{Hawking:1971ei, Carr:1974nx}.  This is because an extra e-folding is required in the second inflation or in the ultra-slow-roll (USR) phase~\cite{Inoue:2001zt, Tsamis:2003px, Kinney:2005vj} responsible for the PBH formation~\cite{Ezquiaga:2017fvi, Motohashi:2017kbs, Germani:2017bcs} so that the e-folding for the first inflation is reduced (see, e.g., Refs.~\cite{Iacconi:2021ltm, Kallosh:2022ggf, Braglia:2022phb}).  To match the spectral index, its dependence on the e-folding number must be compensated. 

%%%%%%%%%%%%%%%%%%%%%%%%%%%%%%%%%%%%
\begin{table}[tbhp]
    \centering
    \caption{\label{tab:relation} Relations to other generalizations of the inflationary $\alpha$-attractor models.}
     \begin{tabular}{|c || c|c|c|} 
     \hline
     pole order & regular potential & singular derivative & singular potential \\
     \hline\hline
     second order & $\alpha$-attractor~\cite{Ellis:2013nxa, Ferrara:2013rsa, Kallosh:2013yoa, Kallosh:2014rga, Carrasco:2015pla, Roest:2015qya, Linde:2015uga, Scalisi:2015qga} & \begin{tabular}{c}polynomial\\ $\alpha$-attractor~\cite{Kallosh:2022feu}\end{tabular} & generalized pole inflation~\cite{Terada:2016nqg} \\ 
     \hline
    general & pole inflation~\cite{Galante:2014ifa, Broy:2015qna} & this work & generalized pole inflation~\cite{Terada:2016nqg} \\
     \hline
     \end{tabular}
\end{table}
%%%%%%%%%%%%%%%%%%%%%%%%%%%%%%%%%%%%

In this work, we investigate the further generalization of such polynomial $\alpha$-attractor with the orders of the pole other than two and explore the inflationary predictions.
Note that the above potential before canonical normalization already has a plateau at the large-field limit. This implies that the proposed potential may lead to a viable inflationary prediction even without the canonical normalization effect of the $\alpha$-attractor.  On the other hand, 
in a generalization of the polynomial $\alpha$-attractor with lower- or higher-order poles, the canonical normalization $\rho \sim \phi^q$ ($q$: const.) does not alter the functional form of the potential significantly
because the logarithmic potential leads to $\ln \rho \sim \ln \phi^q \propto \ln \phi$.
This is a rather exceptional feature in the context of attractor models of inflation, allowing us to study the logarithmic potential of the plateau type with and without the kinetic-term pole in a unified way.

Our work can also be seen as a generalization of pole inflation with the singularity
in the derivative of the potential. The relations of our models to other generalizations of $\alpha$-attractor models are summarized in Table~\ref{tab:relation}. We aim to fill the gap in the literature and to
achieve a more comprehensive understanding of the role of singularities in the kinetic term
and in the potential term.

The present paper is organized as follows.
In Section~\ref{sec:model}, we introduce the inflation model with logarithmic potentials and an arbitrary order of the pole in the kinetic term. We study the predictions of the models in Section~\ref{sec:predictions}. As an application, we study the PBH formation and the associated (curvature-induced) gravitational-wave production with a slightly modified logarithmic potential in Section~\ref{sec:pbh}. We find a benchmark point in the parameter space where (i) the whole dark matter abundance is explained by PBHs, (ii) the stochastic gravitational-wave background is testable by LISA, (iii) the spectral index $n_s \approx 0.96$ is consistent with the CMB constraints despite the decrease typical in PBH scenarios, and (iv) the tensor-to-scalar ratio $r \approx 0.025$ is large enough to be tested in the near future. We conclude in Section~\ref{sec:summary}.  We take the natural unit $c = \hbar = 8\pi G = 1$.

%%%%%%%%%%%%%%%%%%%%%%%%%%%%%%%%%%%%%%%%%%%%
\section{Pole inflation
with singular derivative of potential \label{sec:model}}
%%%%%%%%%%%%%%%%%%%%%%%%%%%%%%%%%%%%%%%%%%%%

We will briefly review the polynomial $\alpha$-attractors~\cite{Kallosh:2022feu} first and then consider its generalization 
with arbitrary orders of poles.
In the polynomial $\alpha$-attractor model, the Lagrangian density is constructed by 
 the kinetic term with a second-order pole~\footnote{
The tilde on $\tilde\alpha$ is put to distinguish it from $\alpha$ which is introduced below in Eq.~\eqref{eq:logarithmicpotential}.  For the second-order pole case ($p=2$), they coincide with each other up to a constant factor, $(3/4)\tilde{\alpha} = \alpha$.
}
\begin{align}
    \mathcal{L}=-\frac{3\tilde \alpha}{4}\frac{(\partial \rho)^2}{\rho^2}-V(\rho), \label{L_polynomial_alpha-attractor}
\end{align}
and a logarithmic potential
\begin{align}
    V(\rho) = V_0 \frac{(\ln \rho)^{2n}}{(\ln \rho)^{2n} + c^{2n}},
    \label{eq:potential_with_n}
\end{align}
where $V_0 $, $c$, and $n$ are constants.
This potential is non-singular, but the derivative is singular at $\rho \to 0$. 

After canonical normalization $\ln \rho = - \sqrt{2/(3\tilde \alpha)} \phi$, the potential of the canonically normalized scalar field
\begin{align}
    V(\phi) = V_0 \frac{\phi^{2n}}{\phi^{2n} + \mu^{2n}},
    \label{eq:nsrinalphaattractor}
\end{align}
where $\mu^{2n} = (3\tilde \alpha/2)c^{2n}$ and its first derivative are not singular.  At large $\phi$, the potential can be expanded as $V \sim V_0 \left(1 - \frac{\mu^{2n}}{\phi^{2n}} + \cdots \right)$. Its predictions for $n_s$ and $r$ are given by 
\begin{align}
    n_{s} =& 1-\frac{2}{N_e}\frac{2n+1}{2n+2}, 
    & r = & \frac{2\left(6 c^{2n} n \tilde{\alpha} \right)^{\frac{1}{1+n}} }{\left((1+n)N_e \right)^{2 - \frac{1}{1+n}}}.
    \label{genns}
\end{align}
The motivation of this generalization in the original context is to explore the possibility of realizing larger values of $n_s$ in the $\alpha$-attractor models since the basic prediction $n_s = 1- 2/N_e$ of the $\alpha$-attractor models is a bit smaller than the central value of the CMB constraints. 

Hereafter, we generalize this model and consider the inflation model in which the order of the pole $p$ is generalized. The Lagrangian density is
\begin{align}
	\mathcal{L} = & -\frac{\alpha}{\rho^p}(\partial \rho)^2 - V(\rho), & V(\rho) = & V_0\frac{(\ln \rho)^2}{(\ln \rho)^2+c^2}\ % \ (p\neq 2)
 .
	\label{eq:logarithmicpotential}
\end{align}
Further generalizations are discussed in the next section. 
Note that the change of variable $\rho = 1/\rho_\text{dual}$ leads to the same Lagrangian density with $p$ replaced with $p_\text{dual} = 4 - p$. 
For $p \neq 2$, the canonical inflaton $\phi$ is given by
\begin{align}
    \phi =
    \frac{2\sqrt{\alpha}}{|p-2|}\rho ^{\frac{2-p}{2}} . \qquad  (p\neq 2) \label{eq:phi-rho_relation}
\end{align}
For both models with $p < 2$ and $p > 2$, the canonical potential is 
\begin{align}\label{eq:canonical-potential}
	V=V_0\left(1-\frac{c^2}{\left(\frac{2}{2-p}\ln\left(\frac{|2-p|}{2\sqrt{\alpha}}\phi\right)\right)^2 + c^2}\right).
\end{align}
Reflecting the above symmetry in terms of $\rho$, the canonical potential is invariant under the change of $p$ to $p_\text{dual}= 4 - p$. For example, the predictions of the model with no kinetic-term pole $p=0$ are the same as those of $p=4$.  Without loss of generality, we focus on $p > 2$. 

The slow-roll parameters $\epsilon_V$ and $\eta_V$ in our model are 
\begin{align}
	\epsilon_V &:= \frac{1}{2}\left(\frac{V'}{V}\right)^2= \frac{8 c^4}{(p-2)^2 \phi^2 (\ln\rho)^2 \left(c^2+(\ln\rho)^2\right)^2},\label{eq:ns}\\
	\eta_V &:= \frac{V''}{V} =\frac{4 c^2 \left(2 c^2 + c^2 (p-2) \ln \rho -6 (\ln\rho)^2+(p-2) (\ln\rho)^3 \right)}{(p-2)^2 \phi^2 (\ln\rho)^2 \left(c^2+(\ln\rho)^2\right)^2},\label{eq:r}
\end{align}
where a prime denotes derivative with respect to $\phi$, and $\rho=\rho(\phi)$ is the inverse of Eq.~\eqref{eq:phi-rho_relation}.
The length of the inflation period, which should be large enough to realize an isotropic and homogeneous universe, is represented by the e-folding number $N_\mathrm{e}$. It is given as follows:
\begin{align}
    N_e &:=\int_t^{t_\mathrm{end}}H \mathrm{d}t
	\simeq -\int^{\phi_\mathrm{end}}_\phi \frac{V}{V'}\mathrm{d}\phi \simeq \frac{\phi^2}{8c^2 (p-2)^2}\Bigl[(p-2)^3(\ln\rho)^3+3(p-2)^2 (\ln\rho)^2 \nonumber\\
    &+((p-2)6+c^2(p-2)^3)\ln\rho +(6+c^2(p-2)^2)\Bigr],  
    \label{eq:analytical_e-foldingNumber}
\end{align}
where $\phi_\mathrm{end}$ is the inflaton field at the end of inflation.
The CMB parameters, namely the scalar spectral index $n_s = 1-6\epsilon_V+2\eta_V$ and the tensor-to-scalar ratio $r= 16 \epsilon_V $ are written in terms of the slow-roll parameters $\epsilon_V$ and $\eta_V$.

It is standard to invert the relation $N_e = N_e(\phi)$~\eqref{eq:analytical_e-foldingNumber} and express the inflationary observables in terms of $N_e$.  For this purpose, let us take the leading term in Eq.~\eqref{eq:analytical_e-foldingNumber} in the limit $\phi\to \infty$.  This leads to $N_\mathrm{e} \simeq  \frac{1}{c^2(p-2)^2} \phi^2\left(\ln \left(\frac{|p-2|}{2\sqrt{\alpha}}\phi \right)\right)^3$.  The $n_s$ and $r$ in this limit are
\begin{align}
	n_s & \simeq 1-\frac{1}{N_\mathrm{e}},& r & \simeq \frac{16c^2(p-2)^2}{N_e(\ln N_e)^3},
\end{align}
up to logarithmic corrections inside the logarithm. 
Comparing these expressions with eq.~\eqref{genns} with $n=1$, we see that our generalization in the large $N_e$ limit has different parametric dependence.  In particular, it predicts larger $n_s$ and larger $r$ than the polynomial $\alpha$-attractor.
However, we stress that the large $N_e$ limit in our model is \emph{not} precise enough at the CMB scale, where the typical $N_e \sim 60$ corresponds to  $\phi \sim \mathcal{O}(10)$ and $\ln \rho \sim \mathcal{O}(1)$. Around this region, the second and third terms in Eq.~\eqref{eq:analytical_e-foldingNumber} are comparable to the first term, which is formally the leading term.  In our numerical calculations, we do not rely on this approximation and instead use expressions like those in Eqs.~\eqref{eq:ns} and \eqref{eq:r}.

%%%%%%%%%%%%%%%%%%%%%%%%%%%%%%%%%%%%%%%%%%%%
\section{Model predictions and CMB constraints 
\label{sec:predictions}}
%%%%%%%%%%%%%%%%%%%%%%%%%%%%%%%%%%%%%%%%%%%%

In this section, we numerically compute the spectral index $n_s$ and the tensor-to-scalar ratio $r$ and study their dependence on the model parameters such as the order of the pole $p$, the residue of the pole  $\alpha$, and the parameter $c$ in the potential.   Then, we compare them with the observational CMB constraints~\cite{Planck:2018jri, BICEP:2021xfz}.

We solve the equation of motion for the inflaton $\phi(N_e)$ using the e-folding number $N_e$ as the time variable,\footnote{
In the previous section, $N_e$ is measured backward from the end of inflation, whereas, in this section, it increases from the beginning of the simulation as a time variable, $\mathrm{d}N_e = H \mathrm{d} t$. 
}
\begin{align}
    \left(\frac{\mathrm{d}^2\phi}{\mathrm{d}N_e^2}\right)+\left(3-\frac{1}{2}\left(\frac{\mathrm{d}\phi}{\mathrm{d}N_e}\right)^2\right)\left(\frac{\mathrm{d}\phi}{\mathrm{d}N_e}+\frac{1}{V}\left(\frac{\mathrm{d}V}{\mathrm{d}\phi}\right)\right)=0.\label{eq:bgeq1}
\end{align}
Assuming that the total e-folding of inflation is longer than the observable range, the dynamics approaches that of the slow-roll attractor solution.  In this case, we can neglect $|\mathrm{d}^2\phi/\mathrm{d}N_e^2|$ and $|\mathrm{d}\phi/\mathrm{d}N_e|^2$ in the above equation. Thus, the initial velocity for our simulation is given by 
\begin{align}
    \left. \frac{\mathrm{d}\phi}{\mathrm{d}N_e}\right|_\text{ini}\simeq-\sqrt{2\epsilon_V(\phi_\text{ini})}.
\end{align}
As usual, we take the condition $\epsilon_H=1$ as the end of inflation where $\epsilon_H$ is defined by
\begin{align}
    \epsilon_H(N_e):=& - \frac{1}{H}\frac{\mathrm{d}H}{\mathrm{d}N_e}  = \frac{1}{2}\left(\frac{\mathrm{d}\phi}{\mathrm{d}N_e}\right)^2,
\end{align}
where $H$ is the Hubble variable. 
Solving Eq.~\eqref{eq:bgeq1} for various initial values until the end of inflation, we find the initial values for which $N_e(\phi_\text{end})=50$ and $60$. Then, we can obtain the spectral index $n_s$ and the tensor-to-scalar ratio $r$ from Eqs.~\eqref{eq:ns} and \eqref{eq:r}.

%%%%%%%%%%%%%%%%%%%%%%%%%%%%%%%%%%%%
\begin{figure}[t]
  \centering
  \includegraphics[width=13cm]{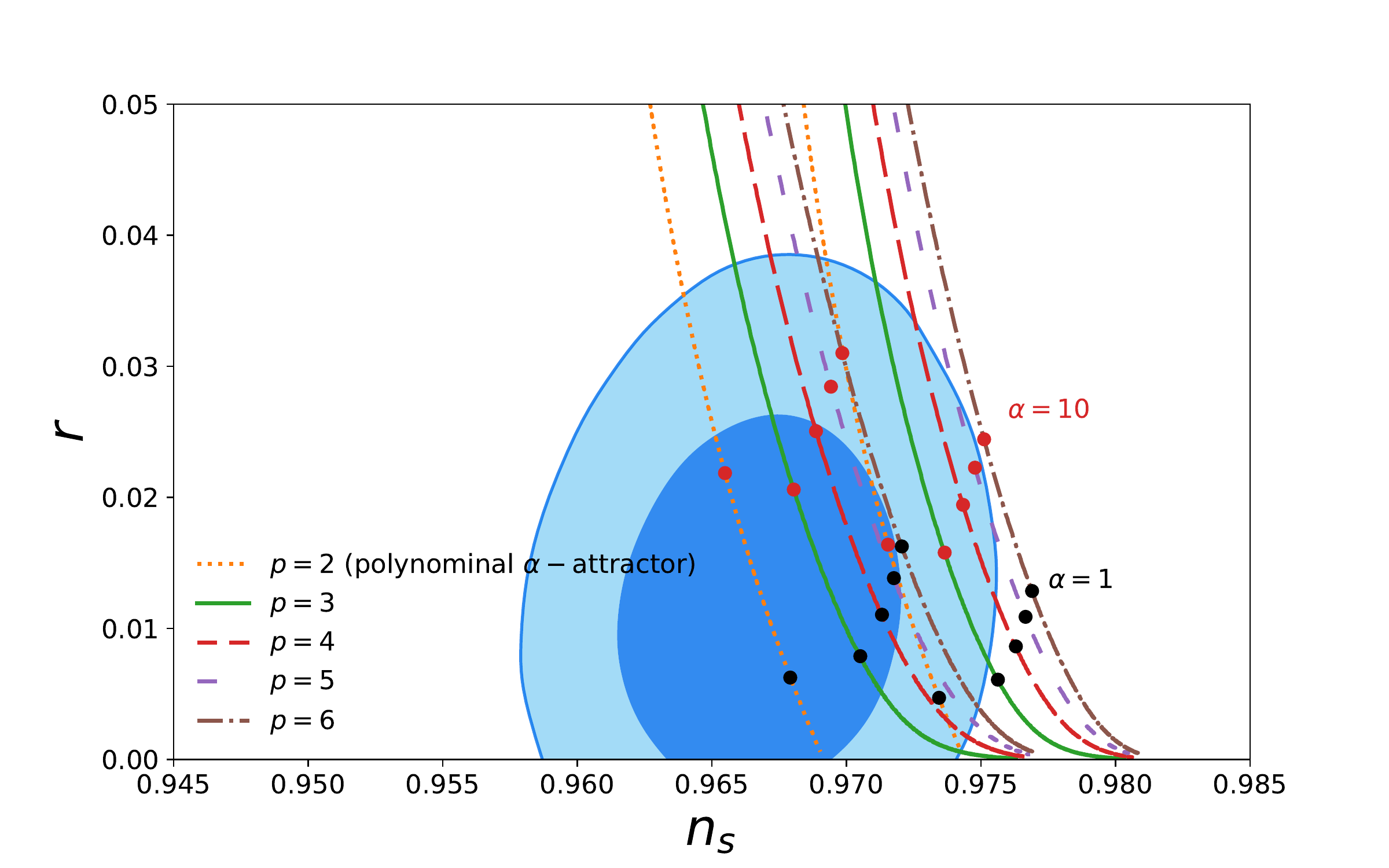}
  \caption{Dependence of the spectral index $n_s$ and the tensor-to-scalar ratio $r$ on the residue $\alpha$ in the model~\eqref{eq:canonical-potential}. The positions of $\alpha = 1$ and $10$ are shown by the black and red point markers, respectively.  The parameter in the potential is set to $c=1$. Each line corresponds to the order of the pole $p=2, 3, 4, 5$, and $6$ (see figure legends). The prediction is the same for $p_\text{dual}=4-p$. The left and right sets of lines correspond to $N_e(\phi_\text{end})=50$ and $60$, respectively.  The CMB constraints on $n_s$ and $r_{0.002}$ at $0.002 \ \mathrm{Mpc}^{-1}$ at 68\% and 95\% confidence levels are shown with the blue contours, drawn by using the code on the web page of BICEP/Keck~\cite{bicepkeck}.}
  \label{fg:nsr1}
\end{figure}
%%%%%%%%%%%%%%%%%%%%%%%%%%%%%%%%%%%%
\begin{figure}[t]
  \centering
  \includegraphics[width=13cm]{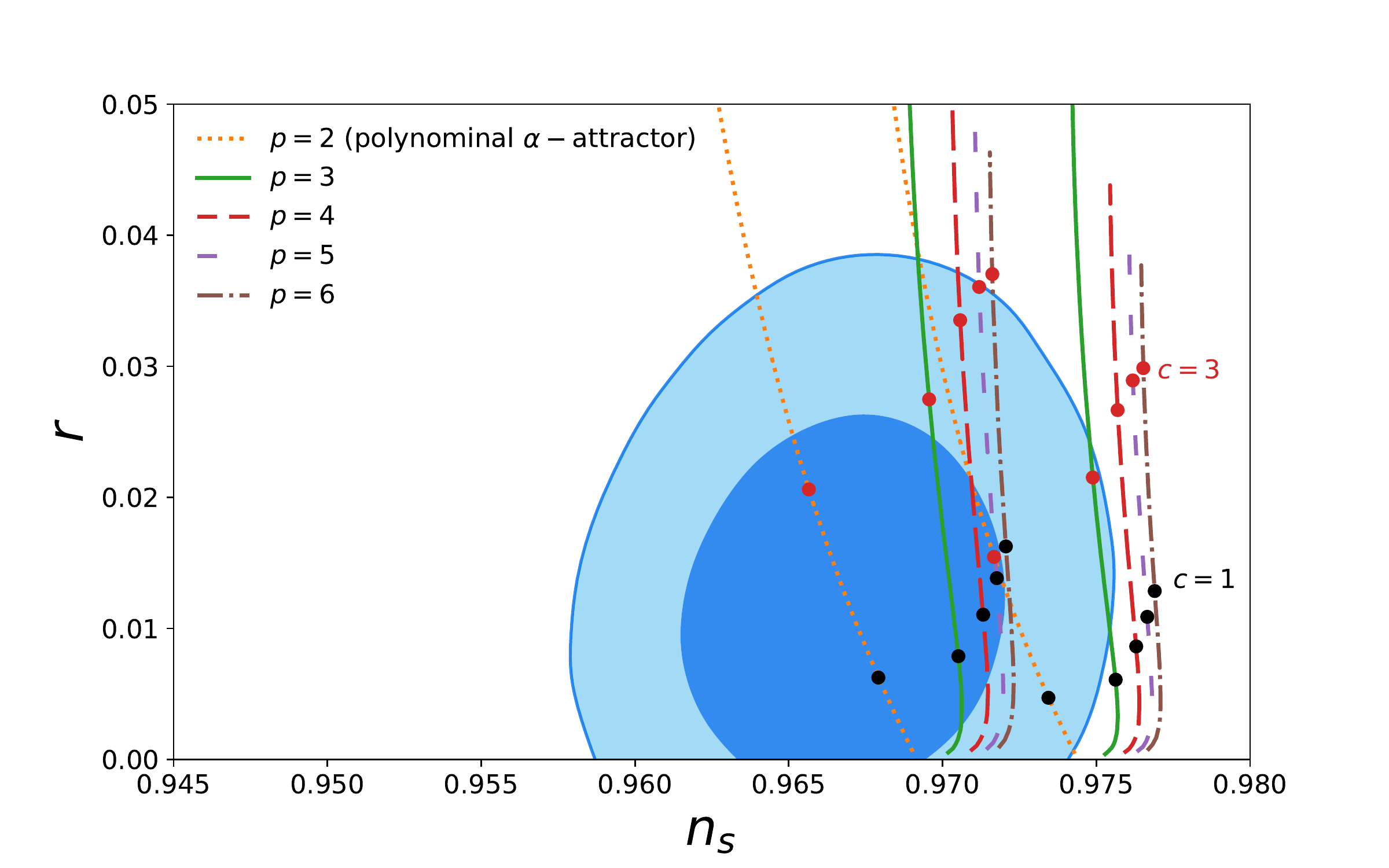}
  \caption{Dependence of the spectral index $n_s$ and the tensor-to-scalar ratio $r$ on the parameter $c$ in the model~\eqref{eq:canonical-potential}. The positions of $c=1$ and $3$ are shown by the black and red point markers, respectively. The residue parameter is set to $\alpha=1$. The rest is the same as in Fig.~\ref{fg:nsr1}.}
  \label{fg:nsr2}
\end{figure}
%%%%%%%%%%%%%%%%%%%%%%%%%%%%%%%%%%%%

We show the relation between $n_s$ and $r$ in the model~\eqref{eq:canonical-potential} in Fig.~\ref{fg:nsr1} and Fig.~\ref{fg:nsr2}. In Fig.~\ref{fg:nsr1}, we plot the dependence on $\alpha$ with $c=1$ fixed for each pole order $p =2, 3, 4, 5$, and $6$, whereas, in Fig.~\ref{fg:nsr2}, we plot the dependence on $c$ with $\alpha=1$ fixed for each pole order. Fig.~\ref{fg:nsr1} shows the models with $p \neq 2$ approach to a larger value of $n_s$ for small $\alpha$, which is different behavior from the original polynomial $\alpha$-attractor $p = 2$. On the other hand, the models approach $n_s \sim 0.970-0.975$ for small values of $c$ and are consistent with observations, which is the same behavior with the original polynomial $\alpha$-attractor. 

The typical values of $n_s$ are a bit larger than the original (exponential) $\alpha$-attractor models. Remember that the motivation of polynomial $\alpha$-attractor was to make $n_s$ larger.
Fig.~\ref{fg:nsr1} and Fig.~\ref{fg:nsr2} show that $n_s$ can be further larger by increasing (decreasing) the pole order for $p>2$ ($p<2$). 
Although the second-order pole ($p=2$) case~\cite{Kallosh:2022feu} is closest to the center of the constraints, we find that a substantial part of the parameter space for $p\neq2$ is also consistent with the observations.  It is interesting to note that even the model without the pole $p=0$ (equivalent to $p=4$) has a sizable allowed region. 
Since the tensor-to-scalar ratio $r$ is relatively large unless $c \ll 1$, this model is readily verifiable in future observations.

Although a large $n_s$ can be an advantage for some cosmological scenarios as discussed in more detail in the next section, it can also be a disadvantage in other scenarios since it is less close to the sweet spot of the contour of $n_s$ and $r$. Therefore, we also discuss some extensions of the model so that $n_s$ is close to the central value of the constraints. 

\begin{figure}[tbh]
  \centering
  \includegraphics[width=13cm]{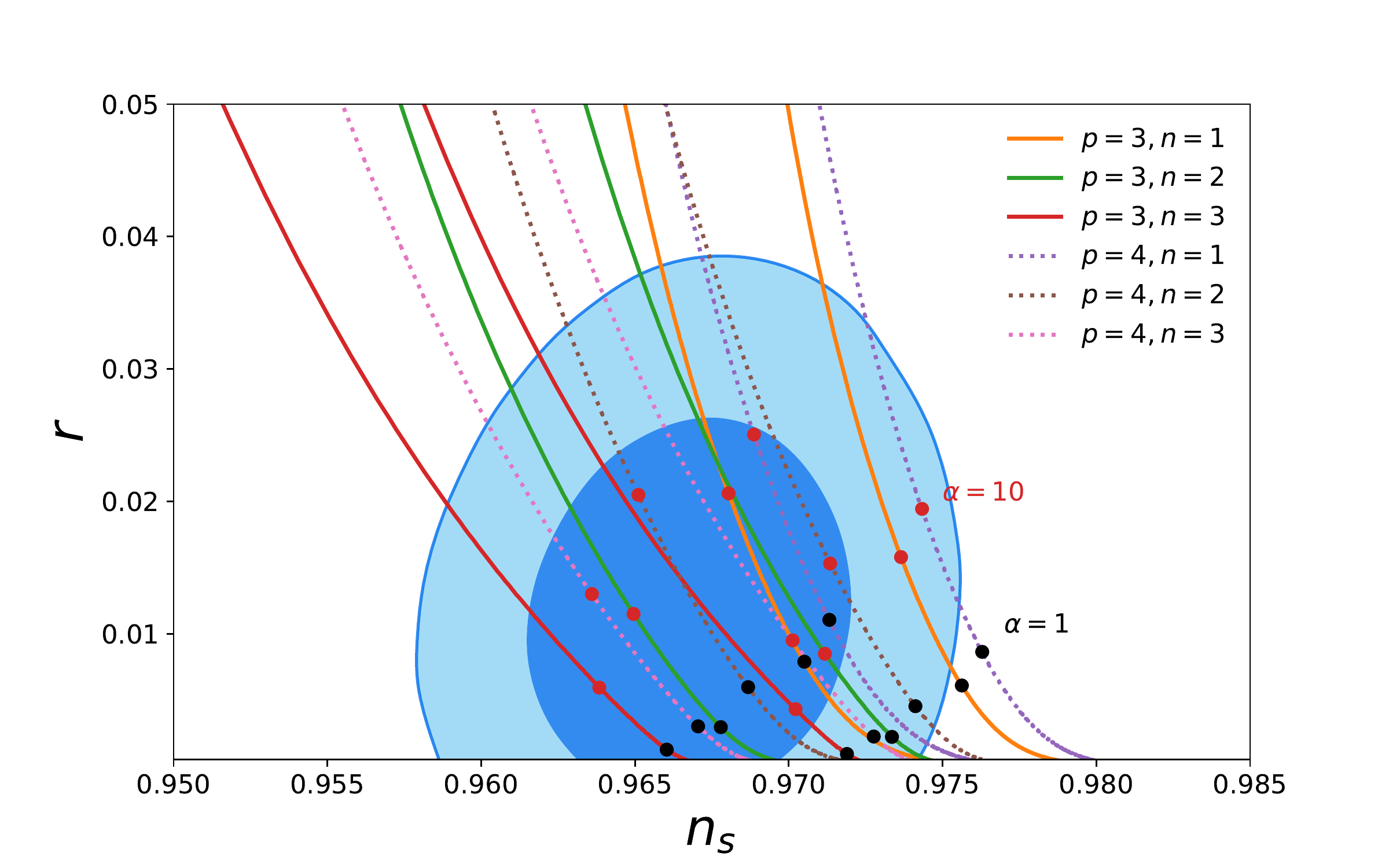}
  \caption{Dependence of the spectral index $n_s$ and the tensor-to-scalar ratio $r$ on the residue $\alpha$ in the model~\eqref{eq:potential_generalize_order_log}.  For different lines, we vary the order of the pole ($p = 3$ and $4$) and the power in the potential ($n = 1, 2$, and $3$) (see the legends) while fixing $c=1$.  The rest is the same as in Fig.~\ref{fg:nsr1}.}
  \label{fg:nsr_generalize_order_log}
\end{figure}

The first extension is obtained by letting the power of $\ln \rho$ in the potential a free parameter $2n$ [Eq.~\eqref{eq:potential_with_n}].  For convenience, we show it again
\begin{align}
    V(\rho)=V_0\frac{(\ln \rho)^{2n}}{(\ln \rho)^{2n}+c^{2n}}.
    \label{eq:potential_generalize_order_log}
\end{align}
This extension was studied for the second-order pole case ($p=2$) in Ref.~\cite{Kallosh:2022feu}. The above potential reduces to the previous  potential~\eqref{eq:logarithmicpotential} when $n = 1$.
We numerically compute the inflationary observables in this model and show the effect of varying the power $n$ of $\ln \rho$ in Fig.~\ref{fg:nsr_generalize_order_log}. 
It shows that $n_s$ is smaller for larger $n$.
This is similar to the $p=2$ case~\cite{Kallosh:2022feu}.

\begin{figure}[tbh]
  \centering
  \includegraphics[width=13cm]{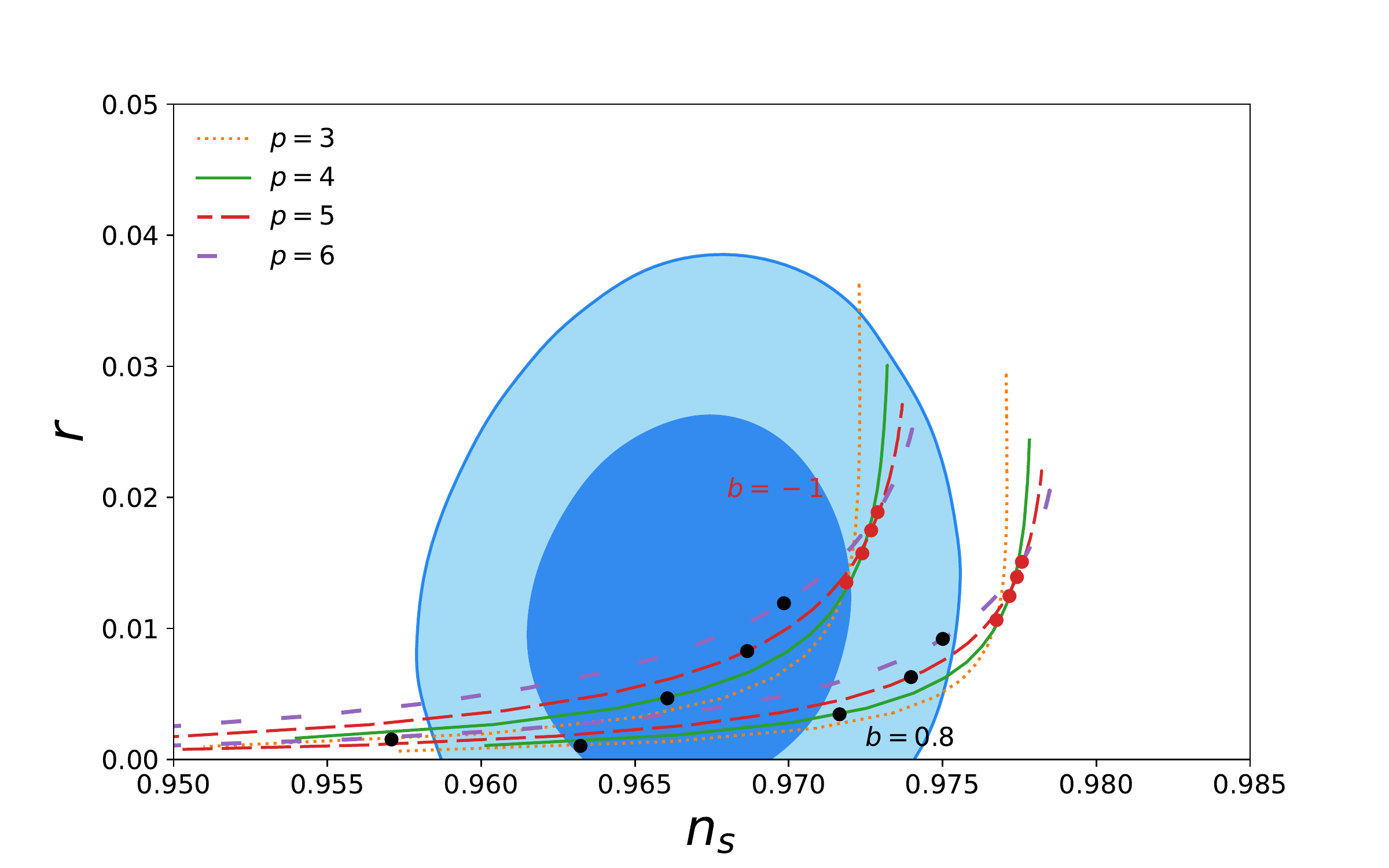}
  \caption{Dependence of the spectral index $n_s$ and the tensor-to-scalar ratio $r$ on the parameter $b$ in the model~\eqref{eq:potential_first_order_log}. The positions of $b=-1$ and $0.8$ are shown by the red and black point markers, respectively. The other parameters are set to $\alpha=1$ and $c=1$. Each line corresponds to the order of the pole $p=3, 4, 5$, and $6$ (see the legends). The prediction is the same for $p_\text{dual}=4-p$. The rest is the same as in Fig.~\ref{fg:nsr1}.
  }
  \label{fg:nsr_first_order_log}
\end{figure}

Another way to reduce $n_s$ is to add the first power of $\ln \rho$ to the denominator of the potential
\begin{align}
    V(\rho)=V_0\frac{(\ln \rho)^2}{b(\ln \rho)+(\ln \rho)^2+c^2}.
    \label{eq:potential_first_order_log}
\end{align}
When we take $b=0$, it reduces to the original potential~\eqref{eq:logarithmicpotential}. 
We show the dependence of the inflationary observables on $b$ in Fig.~\ref{fg:nsr_first_order_log}. 
This figure indicates that the model with a larger positive $b$ predicts smaller $n_s$.

%%%%%%%%%%%%%%%%%%%%%%%%%%%%%%%%%%%%
%%%%%%%%%%%%%%%%%%%%%%%%%%%%%%%%%%%%
%%%%%%%%%%%%%%%%%%%%%%%%%%%%%%%%%%%%

Note that the predictions of the model changed significantly by the above two extensions. This is in contrast to the standard $\alpha$-attractor mechanism, in which the inflationary observables are not sensitive to the details of the original potential. The difference arises from the fact that the canonical normalization for $p\neq 2$ involves a power of the inflaton while its dependence in the potential is logarithmic. Because of this, the attractor nature of our generalization of polynomial $\alpha$-attractor is quite weak. This explains the flexibility of our models.

Before closing the section, let us compare our generalizations $(p\neq 2)$ with the polynomial $\alpha$-attractor ($p=2$) in the large $N_e$ limit. We have done this comparison for the simplest case at the end of Section~\ref{sec:model}, and here we assume the potential in Eq.~\eqref{eq:potential_generalize_order_log} without restricting to $n=1$.   
In the large $N_e$ limit, the e-folding number can be approximated as $N_e \simeq \frac{2^{2n} }{4n(2-p)^{2n}c^{2n}}\phi^2 \left(\ln\left(\frac{(p-2)\phi}{(2\sqrt{\alpha}}\right)\right)^{2n+1}$. The slow-roll parameters are expressed as $\epsilon_V \simeq \frac{nc^{2n}(p-2)^{2n}}{N_e(\ln N_e)^{2n+1}}$ and $\eta_V \simeq \frac{1}{2N_e}$. This yields
\begin{align}
	n_s & \simeq 1-\frac{1}{N_\mathrm{e}},& r & \simeq \frac{16 n c^{2n}(p-2)^{2n}}{N_e(\ln N_e)^{2n+1}}.
\end{align}
Both $n_s$ and $r$ in our generalization are larger than those in the polynomial $\alpha$-attractor [Eq.~\eqref{genns}]. In our case, the $n$ dependence does not show up in the leading term of $n_s$ in the large $N_e$ limit.

%%%%%%%%%%%%%%%%%%%%%%%%%%%%%%%%%%%%%%%%%%%%
\section{Primordial black holes and curvature-induced gravitational waves\label{sec:pbh}}
%%%%%%%%%%%%%%%%%%%%%%%%%%%%%%%%%%%%%%%%%%%%
As discussed in the previous section, inflation models with the logarithmic potential~\eqref{eq:logarithmicpotential} tend to have large $n_s$. In contrast, $n_s$ tends to be small in typical PBH scenarios (except for, e.g., recent work~\cite{Aldabergenov:2023yrk,Wu:2021zta}) in which a substantial amount of PBHs are produced in the post-inflationary universe due to the enhanced primordial curvature perturbations generated by non-attractor dynamics during inflation such as a transition to a temporary USR regime.\footnote{\label{fn:quantum_correction}
Quantum corrections to the power spectrum of curvature perturbations in models beyond the slow-roll attractor solution were recently discussed in Refs.~\cite{Kristiano:2022maq, Inomata:2022yte}. 
Ref.~\cite{Kristiano:2022maq} (see also Ref.~\cite{Cheng:2021lif}) considered the transient USR scenario and claimed that the quantum correction to the CMB scale is comparable to the tree-level result, concluding that the PBH production from single-field inflation is ruled out. However, the opposite conclusion was made in Ref.~\cite{Riotto:2023hoz}.  
Ref.~\cite{Inomata:2022yte} found sizable quantum corrections at the same scale as the tree-level enhancement of the curvature perturbations, but the considered mechanism is different from the transient USR period. 
} A non-negligible amount of inflation happens during the USR period, so the required amount of e-folding prior to the USR period is shortened. Because of this, if the inflaton potentials responsible for PBH formation are obtained by deforming some base inflation model, the $n_s$ and $r$ at the CMB scale after deformation needs to be evaluated at an effectively smaller value of $N_e$ than that in the base model.  Since $n_s - 1$ and $r$ are generically suppressed by some powers of $N_e$, $n_s$ and $r$ in the deformed model become smaller and larger, respectively. Therefore, the large $n_s$ in the logarithmic potential model~\eqref{eq:logarithmicpotential} has the advantage in the PBH scenario as a base inflation model. 

Motivated by this discussion, we deform the logarithmic potential to produce PBHs. We study if the PBHs can play the role of dark matter while $r$ and particularly $n_s$ are kept consistent with CMB constraints. The enhanced primordial curvature perturbations also produce gravitational waves due to the non-linear interactions in General Relativity, so we also study the curvature-induced gravitational waves in this section. 

Inspired by the critical Higgs inflation model for PBH formation~\cite{Ezquiaga:2017fvi}, we consider a deformed logarithmic potential to have an approximate inflection point that enhances curvature perturbations 
\begin{align}
    V(\phi)=&V_0\frac{x^2\left(1+d_1 \left(\ln(x^2)\right)^2\right)}{x^2\left(1+d_2 \left(\ln(x^2)\right)^2\right)+c^2},& x:=&\ln\left(\frac{\phi^2}{\alpha}\right)=2\ln\left(\frac{\phi}{\sqrt{\alpha}}\right).
    \label{eq:logarithmicpotentialwithinflection} 
\end{align}
When the new parameters $d_1$ and $d_2$ vanish, this model reduces to the base model~\eqref{eq:logarithmicpotential} after redefining the parameters.

%%%%%%%%%%%%%%%%%%%%%%%%%%%%%%%%%%%%
\begin{figure}[t]
    \centering
    \includegraphics[width=9cm]{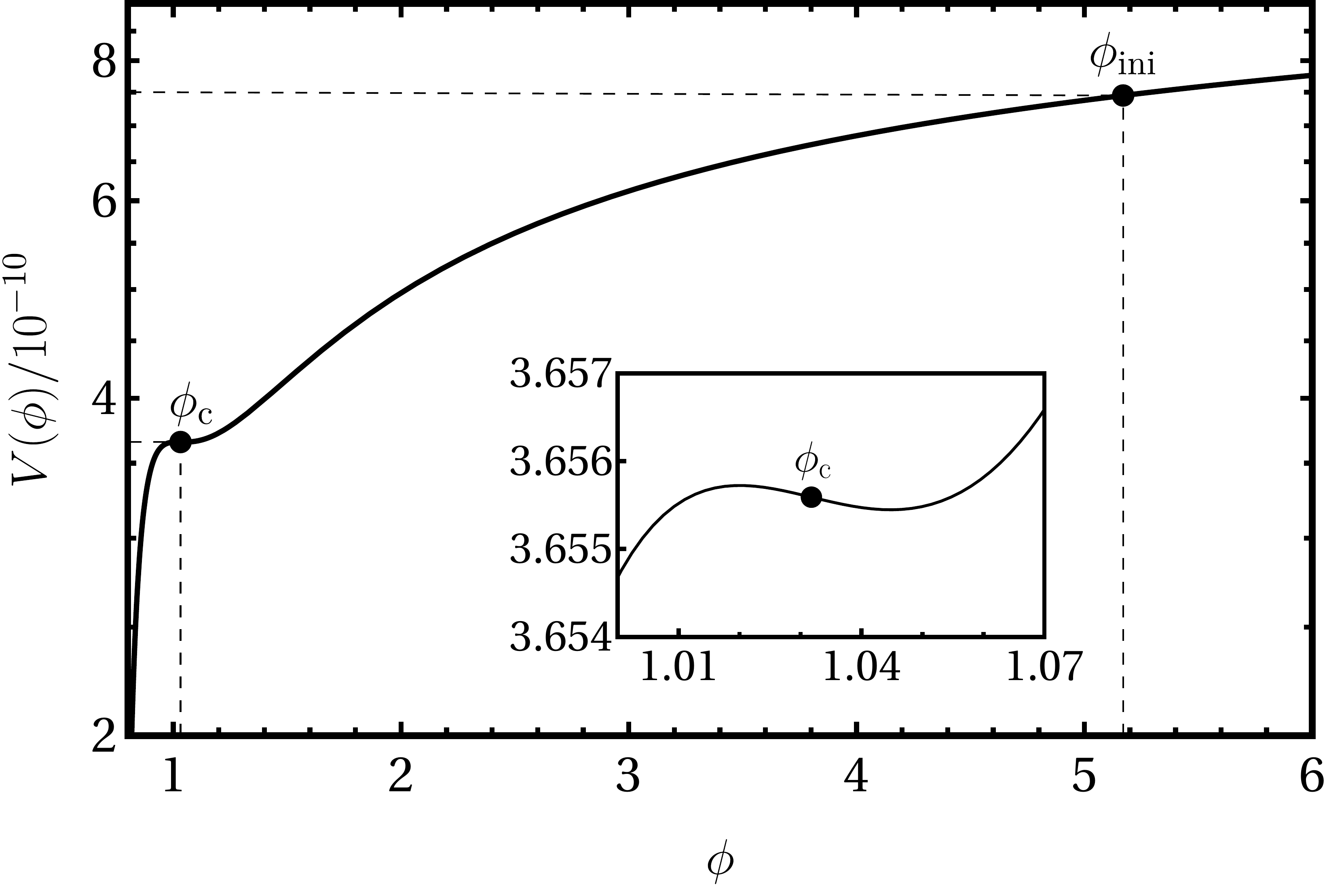}
    \caption{The logarithmic potential modified to have an approximate inflection point~\eqref{eq:logarithmicpotentialwithinflection}.}
    \label{fg:PBHpotential}
\end{figure}
%%%%%%%%%%%%%%%%%%%%%%%%%%%%%%%%%%%%

\subsection{Power spectrum of curvature perturbations}
To estimate the abundance of PBHs and the energy density of the curvature-induced gravitational waves, the power spectrum of curvature perturbations at the end of inflation has to be calculated. In the slow-roll approximation, the power spectrum of curvature perturbations is derived analytically in Ref. \cite{Stewart:1993bc} as
\begin{align}
    \mathcal{P}_\mathcal{R}\simeq \left. \frac{H^2}{2\pi^2}\left(\frac{\mathrm{d}\phi}{\mathrm{d}N_e}\right)^{-2}\right|_{k=aH} \simeq \left. \frac{1}{8\pi^2}\frac{H^2}{\epsilon_V}\right|_{k=aH}.
\end{align}
This equation indicates that the curvature perturbations which cross the horizon when the inflaton reaches near the approximate inflection point $\phi_c$ are enhanced because of $\epsilon_V(\phi_c)\simeq 0$. 
The right-hand side diverges at the point satisfying $V'=0$, but in fact $\mathcal{P}_\mathcal{R}$ does not diverge because the slow-roll approximation breaks down and the inflaton has nonzero velocity~\cite{Motohashi:2017kbs,Germani:2017bcs}. 
To calculate the power spectrum accurately, we use the Mukhanov-Sasaki formalism~\cite{Mukhanov:1988jd,Sasaki:1986hm} without the slow-roll approximation.

We numerically solve the equations of motion for both the background and perturbation modes to search for parameters leading to an appropriate $n_s$ and the PBH abundance for dark matter, following the procedure in Ref.~\cite{Ballesteros:2017fsr}. See also Ref.~\cite{Drees:2019xpp} for detailed discussions.  
The equations of motion for background variables are Eq.~\eqref{eq:bgeq1} and the Einstein equations for flat Friedmann-Lema\^itre-Robertson-Waler background spacetime
\begin{align}
    3H^2=&\frac{1}{2}H^2\left(\frac{\mathrm{d}\phi}{\mathrm{d}N_e}\right)^2+V,\label{eq:bgeq2}\\
    \frac{\mathrm{d}H}{\mathrm{d}N_e}=&-\frac{1}{2}H\left(\frac{\mathrm{d}\phi}{\mathrm{d}N_e}\right)^2. \label{eq:bgeq3}
\end{align}
Two of the background equations~\eqref{eq:bgeq1}, \eqref{eq:bgeq2}, and \eqref{eq:bgeq3} are independent. We numerically solve   Eqs.~\eqref{eq:bgeq1} and \eqref{eq:bgeq2}.
\par 
The equation of the comoving curvature perturbation $\mathcal{R}_k$ for each $k$ mode, which is known as the Mukhanov-Sasaki equation~\cite{Mukhanov:1988jd,Sasaki:1986hm}, is given by
\begin{align}
    \frac{\mathrm{d}^2u_k}{\mathrm{d}N_e^2}+(1-\epsilon_H)\frac{\mathrm{d}u_k}{\mathrm{d}N_e}+\left(\left(\frac{k}{aH}\right)^2+(1+\epsilon_H-\eta_H)(\eta_H-2)-\frac{\mathrm{d}(\epsilon_H-\eta_H)}{\mathrm{d}N_e}\right)u_k=0,\label{eq:MSeq}
\end{align}
where
\begin{align}
    u_k(N_e)=&-z\mathcal{R}_k,& &  & 
    z(N_e)=&a\frac{\mathrm{d}\phi}{\mathrm{d}N_e}, \nonumber \\
    \epsilon_H(N_e)=&\frac{1}{2}\left(\frac{\mathrm{d}\phi}{\mathrm{d}N_e}\right)^2, & &\text{and} & 
    \eta_H(N_e) :=& - \frac{\ddot{\phi}}{H \dot{\phi}} = \epsilon_H-\frac{1}{2}\frac{\mathrm{d}\ln\epsilon_H}{\mathrm{d}N_e}.
\end{align}
We take the Bunch-Davies vacuum for sub-horizon scales $k \gg aH$ as the initial condition. Then, the function $u_k$ for $\eta\to -\infty$ becomes
\begin{align}
    u_k\to\frac{e^{-ik\eta}}{\sqrt{2k}},
\end{align}
where $\eta$ is the conformal time which is related to time by $\mathrm{d}t=a\mathrm{d}\eta$. For numerical calculations, it is convenient to solve the real and imaginary parts of the Mukhanov-Sasaki equation~\eqref{eq:MSeq} separately. We take the initial condition of $u_k$ as
\begin{align}
    \left. \Re (u_k)\right|_\text{ini}=&\frac{1}{\sqrt{2k}},&
    \left. \Im (u_k)\right|_\text{ini}=&0,&
    \left. \Re \left(\frac{\mathrm{d}u_k}{\mathrm{d}N_e}\right)\right|_\text{ini}=&0,&
    &\text{and}&
    \left. \Im \left(\frac{\mathrm{d}u_k}{\mathrm{d}N_e}\right)\right|_\text{ini}=&-\frac{\sqrt{k}}{\sqrt{2}k_i},&
\end{align}
where $k_i \ll k$ for each $k$, taken to be the horizon scale 3 e-folds before the time $k$ crosses the horizon in our calculation. At the sufficiently late time after the horizon crossing, the power spectrum of the curvature perturbation is given by
\begin{align}
    \mathcal{P}_\mathcal{R}=\frac{k^3}{2\pi^2}\left|\frac{u_k}{z}\right|^2_{k\ll aH}.
\end{align}
\par
\begin{figure}[t]
    \begin{minipage}[t]{0.48\linewidth}
        \centering
        \includegraphics[width=1.0\linewidth]{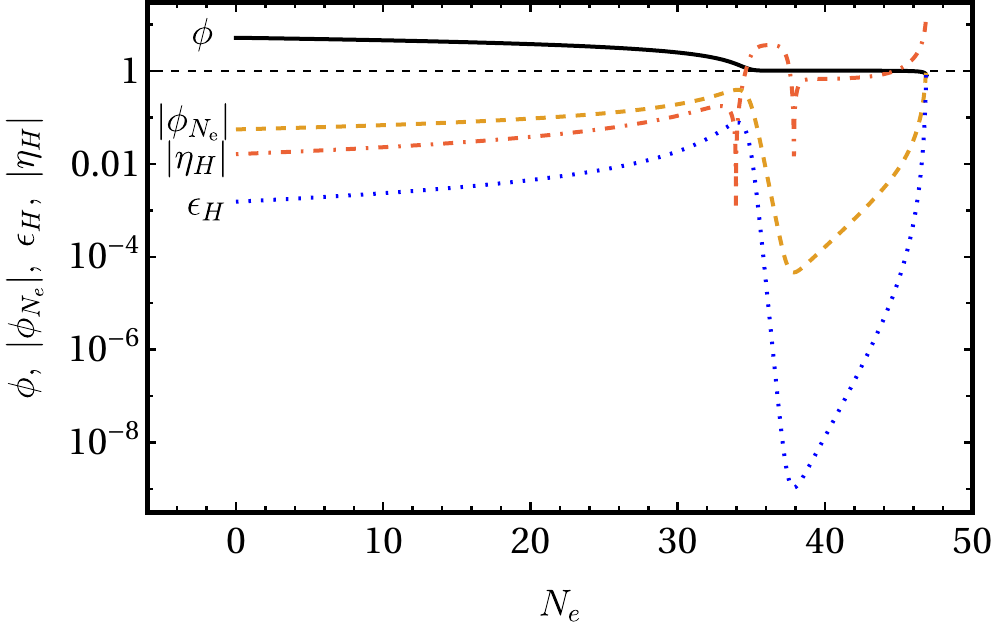}
        \caption{Time evolution of $\phi$, $|\phi_{N_e}|$, $\epsilon_H$ and $|\eta_H|$.}
        \label{fg:timeevolution}
    \end{minipage}
    \begin{minipage}[t]{0.04\linewidth}
    \end{minipage}
    \begin{minipage}[t]{0.48\linewidth}
        \centering
        \includegraphics[width=1.0\linewidth]{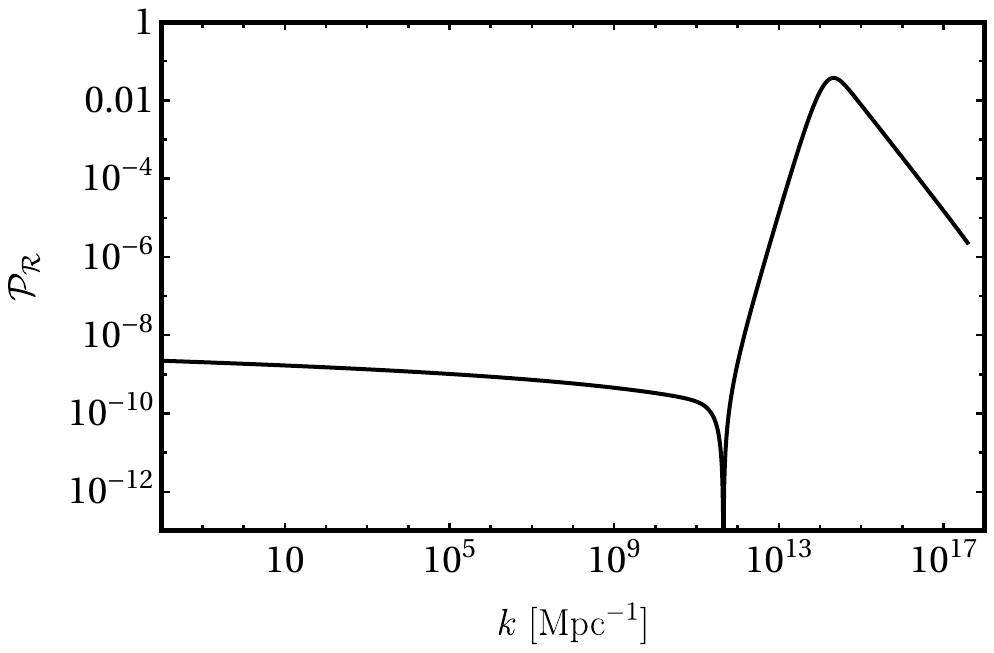}
        \caption{Power spectrum of the curvature perturbations}
        \label{fg:powerspectrum}
    \end{minipage}
\end{figure}

We find that the following parameters of the potential~\eqref{eq:logarithmicpotentialwithinflection},
\begin{align}
    V_0 =& 4.2110582 \times 10^{-10},& \alpha =& 0.58106463,\nonumber \\
    c =& 0.33489191,& d_1 =& 0.15610108 ,&d_2 =& 0.026912140,
    \label{eq:parameters}
\end{align}
and the initial value $\phi_\text{ini}=5.22$ lead to the results
\begin{align}
    n_s =& 0.961,& r=& 0.0249, &\left. N_e\right|_\text{end} =& 46.9, & \phi_\text{end} =& 0.825,
\end{align}
and
\begin{align}
    \max \mathcal{P}_\mathcal{R}(k) =0.0375 \qquad  &\text{ at }k=2.16 \times 10^{14} \ \mathrm{Mpc^{-1}},\\
    A_\text{s}=\mathcal{P}_\mathcal{R}(k) =2.10 \times 10^{-9} \qquad  &\text{ at }k=0.05\ \mathrm{Mpc^{-1}}.
\end{align}
Fig.~\ref{fg:timeevolution} shows the time evolution of $\phi$, $|\phi_{N_e}|:= |\mathrm{d}\phi/\mathrm{d}N_e|$,  $\epsilon_H$, and $|\eta_H|$. Since $\eta_H$ becomes $\mathcal{O}(1)$ at the beginning of the USR region, the slow-roll approximation is not valid. This is why we have to solve the Mukhanov-Sasaki equation.  Fig.~\ref{fg:powerspectrum} shows the power spectrum of curvature perturbations. Since $\epsilon_H$ gets also close to $0.1$ at the beginning of the USR region $N_e\sim 34$, the friction term of the Mukhanov-Sasaki equation~\eqref{eq:MSeq} becomes smaller and the curvature perturbation with the horizon scale $k \sim 10^{14}\ \mathrm{ Mpc^{-1}}$ grows larger. For an analytic understanding of the steep growth, see Refs.~\cite{Byrnes:2018txb, Carrilho:2019oqg, Cole:2022xqc, Liu:2020oqe}.  This large curvature perturbation collapses into PBHs and generates large energy-density gravitational waves when the mode reenters the horizon after inflation.

%%%%%%%%%%%%%%%%%%%%%%%%%%%%%%%%%%%%
%%%%%%%%%%%%%%%%%%%%%%%%%%%%%%%%%%%%

\subsection{Primordial black holes}
Using the power spectrum of the primordial curvature perturbations obtained in the previous subsection with the parameters~\eqref{eq:parameters}, we calculate the abundance of PBHs, following Ref.~\cite{Ballesteros:2017fsr}. 
We use Carr's formula, or the Press-Schechter formalism, to estimate the abundance of the PBHs~\cite{Carr:1975qj}. There are more advanced calculation techniques like those based on the peaks theory~\cite{1986ApJ...304...15B, Yoo:2018kvb, Germani:2018jgr, Yoo:2020dkz} and extra physical effects like the critical collapse~\cite{Niemeyer:1997mt, Yokoyama:1998xd, Green:1999xm, Kuhnel:2015vtw} and non-Gaussianity~\cite{DeLuca:2019qsy, Atal:2019cdz, Biagetti:2021eep, Kitajima:2021fpq, Escriva:2022pnz,Pi:2022ysn}, but our purpose here is just proof of principle. For more details on physics of PBHs, see reviews~\cite{Khlopov:2008qy, Sasaki:2018dmp, Carr:2020gox, Carr:2020xqk,Escriva:2022duf}. 

First, we briefly review the derivation of the abundance following Refs.~\cite{Ballesteros:2017fsr, Inomata:2017okj}. 
A region with large density fluctuations above the threshed $\delta\rho/\rho > \delta_c$ partially collapses into a black hole when they reenter the horizon. 
The ratio $\gamma$ between the PBH mass and the mass inside the horizon is analytically estimated to be $\gamma\simeq 3^{-3/2}\simeq 0.2$~\cite{Carr:1975qj}, and we use this value. 
The mass of the PBH produced by the collapse of the overdense region with the length $a/k$ can be written as
\begin{align}
    M(k)=&\left. \gamma\rho\frac{4\pi}{3H^3}\right|_{k=aH}=\gamma M_{H,\text{eq}}\left(\frac{g_*(T_\text{f})}{g_*(T_\text{eq})}\right)^{1/2}\left(\frac{g_{*s}(T_\text{f})}{g_{*s}(T_\text{eq})}\right)^{-2/3}\left(\frac{k}{k_\text{eq}}\right)^{-2}\nonumber\\
    =&10^{18}\text{ g}\left(\frac{\gamma}{0.2}\right)\left(\frac{g_*(T_\mathrm{f})}{106.75}\right)^{1/2}\left(\frac{g_{*s}(T_\mathrm{f})}{106.75}\right)^{-2/3}\left(\frac{k}{7\times 10^{13}\ \mathrm{Mpc^{-1}}}\right)^{-2},
    \label{eq:PBHmass}
\end{align}
where $g_*(T)$ and $g_{*s}(T)$ are the effective relativistic degrees of freedom for energy density and entropy density, $T_\text{eq}$ and $T_\text{f}$ are the temperature at matter-radiation equality and PBH formation, and $M_{H,\text{eq}}$ is the horizon mass at the matter-radiation equality.
The second equality follows from the conservation of entropy $g_{*s}T^3 a^3=\text{const.}$ and the relation between the energy density and temperature in the radiation-dominated era $\rho \propto g_* T^{4}$.

Assuming the density fluctuation $\delta$ satisfies the Gaussian statistics, the probability that the fluctuation that collapses into a PBH with mass $M$ exceeds the threshold $\delta_c \simeq 0.45$~\cite{Musco:2004ak,Polnarev:2006aa} (see also Refs.~\cite{Musco:2008hv,Musco:2012au,Harada:2013epa}), is given by
\begin{align}
    \beta(M)=\frac{1}{\sqrt{2\pi\sigma^2(M)}}\int_{\delta_\text{c}}^\infty \mathrm{d}\delta \exp\left(\frac{-\delta^2}{2\sigma^2(M)}\right)=\frac{1}{2}\text{erfc}\left(\frac{\delta_\text{c}}{\sqrt{2}\sigma(M)}\right) ,
\end{align}
where $\text{erfc}(x)$ is the complementary error function which is defined by $\text{erfc}(x):=\frac{2}{\sqrt{\pi}}\int_x^\infty\mathrm{d}ye^{-y^2}$. 
$\sigma$ denotes the standard deviation of the coarse-grained density contrast, which is defined by
\begin{align}
    \sigma^2(M)=\int \frac{\mathrm{d}q}{q}\mathcal{P}_\delta (q)W(qR)^2=\frac{16}{81}\int \frac{\mathrm{d}q}{q}(qR)^4\mathcal{P}_\mathcal{R} (q)W(qR)^2 ,
\end{align}
where $W(x)$ is a window function, chosen as $W(x)=\exp(-x^2/2)$ in our calculation, and we have used the relation between the density perturbations and the curvature perturbations in the radiation-dominated era.  

The total mass of PBHs with mass $M$ is $\rho_\text{PBH}(M)=\gamma \beta(M) \rho|_{k=aH}$.
The ratio of the energy density of PBHs and dark matter is given by~\cite{Inomata:2017okj,Kohri:2018qtx} 
\begin{align}
    \frac{\Omega_\text{PBH}(M)}{\Omega_\text{DM}}=&\left. \frac{\rho_\text{PBH}}{\rho_\text{matter}}\right|_\text{eq}\frac{\Omega_\text{matter}}{\Omega_\text{DM}}=\frac{g_*(T_\mathrm{f})}{g_*(T_\mathrm{eq})}\frac{g_{*s}(T_\mathrm{eq})}{g_{*s}(T_\mathrm{f})}\frac{T_\text{f}}{T_\text{eq}}\gamma\beta(M)\frac{\Omega_\text{matter}}{\Omega_\text{DM}}\nonumber\\
    =&\frac{\beta(M)}{7 \times 10^{-16}}\left(\frac{\gamma}{0.2}\right)^{3/ 2}\left(\frac{g_*\left(T_\text{f} \right)}{106.75}\right)^{-1 / 4}\left(\frac{g_{*s}\left(T_\text{eq} \right)}{3.909}\right)\left(\frac{g_*\left(T_\text{eq} \right)}{3.363}\right)^{-1}\left(\frac{M}{10^{18} \mathrm{g}}\right)^{-1 / 2},
\end{align}
where $\Omega_X \equiv \rho_X / \rho_\text{total}$ ($X$: PBH, DM, or matter) is the density parameter.
The second equality follows from $g_{*s}T^3a^3=\text{const.}$, $\rho\propto g_{*}T^4$ and $\rho_\text{PBH}\propto a^{-3}$.
The third equality follows from Eq.~\eqref{eq:PBHmass}, Friedmann equation, and $\rho\simeq\rho_\text{r}=\pi^2g_*T^4/30$.
The total abundance of the PBHs is obtained by integrating $\Omega_\text{PBH}(M)$ over the logarithm of the PBH mass 
\begin{align}
    \Omega_\text{PBH}=\int\frac{\mathrm{d}M}{M}\Omega_\text{PBH}(M).
\end{align}
The present abundance of PBHs is conventionally normalized by the dark matter abundance, $f_\text{PBH}(M) \equiv \Omega_\text{PBH}(M)/\Omega_\text{DM}$ and $f_\text{PBH} \equiv \Omega_\text{PBH}/\Omega_\text{DM}$. 

\begin{figure}[t]
  \centering
  \includegraphics[width=12cm]{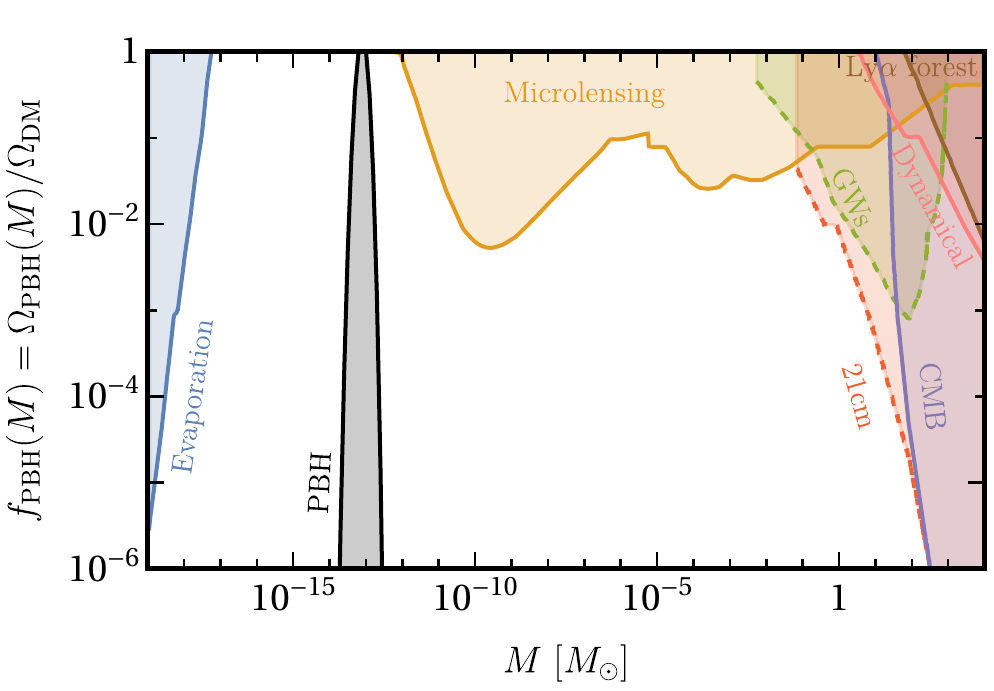}
    \caption{
    Example of the PBH mass distribution in our model (in black) and observational constraints on the PBH abundance (in color). 
    The solid black contour is the abundance of PBH in our model~\eqref{eq:logarithmicpotentialwithinflection} with the  parameters~\eqref{eq:parameters} (same as in Fig.~\ref{fg:powerspectrum}). Constraints are drawn based on Fig.~2 in Ref.~\cite{Villanueva-Domingo:2021spv}. The solid blue contour is the constraint associated with the effect of the PBH evaporation on extra-galactic $\gamma$-ray background~\cite{Carr:2009jm} and CMB~\cite{Clark:2016nst}. The solid orange contour is associated with their microlensing effect on the MACHO~\cite{Macho:2000nvd}, EROS~\cite{EROS-2:2006ryy}, Kepler~\cite{Griest:2013aaa}, Icarus~\cite{Oguri:2017ock}, OGLE~\cite{Niikura:2019kqi}, and Subaru-HSC~\cite{Croon:2020ouk}. The solid purple contour is associated with the accretion effect on the CMB~\cite{Serpico:2020ehh}. The solid pink contour is associated with the dynamical effect on wide binaries~\cite{MonroyRodriguez} and ultra-faint dwarf galaxies~\cite{Brandt:2016aco}. The solid brown contour is associated with the power spectrum of the Lyman-$\alpha$ forest~\cite{Murgia:2019duy}. The dashed green contour is associated with gravitational waves from individual black hole merger events~\cite{Kavanagh:2018ggo, LIGOScientific:2019kan} and the stochastic gravitational wave background~\cite{Chen:2019irf}. This gravitational wave constraint may not be valid~\cite{Boehm:2020jwd}. The dashed red contour is a prospective constraint associated with the 21 cm line~\cite{Mena:2019nhm, Villanueva-Domingo:2021cgh}.}
  \label{fg:PBHAbundance}
\end{figure}

Fig.~\ref{fg:PBHAbundance} shows the abundance of PBHs produced by the approximate inflection point in the potential~\eqref{eq:logarithmicpotentialwithinflection} as a function of its mass by the black line. Various observational constraints are also shown by shaded regions with color. 
PBHs of their masses around $10^{-13}\ M_\odot$ are produced in our model. Two types of constraints were proposed for this mass range.
One of the constraints is associated with the expansion of white dwarfs~\cite{Graham:2015apa} and destruction of neutron stars by PBHs~\cite{Capela:2013yf} which lead to the constraints in the mass range of $10^{-14}-10^{-13}\ M_\odot$ and $10^{-14}-10^{-13}\ M_\odot$, respectively, but the hydrodynamic simulations indicated that this constraint is not effective~\cite{Montero-Camacho:2019jte}.
Another type of constraint comes from null observations of microlensing of stars caused by passing of PBHs~\cite{Niikura:2017zjd} in the line of sight which leads the constraint in the mass range of $10^{-13}-10^{-6}\ M_\odot$, but this constraint below $10^{-11}\ M_\odot$ is not effective as well due to the wave effect of light~\cite{Katz:2018zrn, Inomata:2018cht, gould1992femtolensing}.
Thus, currently, at least the window in the mass range $10^{-14}-10^{-12}\ M_\odot$ remains open (See also Ref.~\cite{ReyIdler:2022unr} and Appendix A in Ref.~\cite{Bartolo:2018rku} for more details).

In the allowed mass range, the total fraction at our benchmark point is~\footnote{The result is presented only in one significant digit as we used values like $\gamma \simeq 0.2$.  However, in our model, we can adjust the maximum value of the power spectrum without significantly changing $n_s$ and $r$ by fine-tuning $\alpha$. Even if the window function, $\gamma$, $\delta_c$, etc. are changed somewhat, we will find the value of $\alpha$ that leads to qualitatively the same results.}
\begin{align}
    f_\text{PBH} =1.
    \label{eq:totalfrction}
\end{align}
Fig.~\ref{fg:PBHAbundance} and Eq.~\eqref{eq:totalfrction} mean that PBHs can play the role of dark matter while avoiding the constraints from current observations.

%%%%%%%%%%%%%%%%%%%%%%%%%%%%%%%%%%%%
%%%%%%%%%%%%%%%%%%%%%%%%%%%%%%%%%%%%

\subsection{Curvature-induced gravitational waves}
In the previous subsection, we discussed that the enhanced primordial curvature perturbations lead to PBHs that can play the whole dark matter. Such enhanced curvature perturbations can also lead to gravitational-wave production through the non-linear interactions in General Relativity~\cite{Saito:2008jc, Saito:2009jt}.  As is well known, the scalar and tensor perturbations are decoupled in the linearized theory, but the latter is produced at the second (or higher~\cite{Yuan:2019udt, Zhou:2021vcw, Chang:2022nzu}) order of cosmological perturbations from the first-order scalar perturbations~\cite{Ananda:2006af, Baumann:2007zm}. Here, we call them curvature-induced gravitational waves, but they are also called second-order gravitational waves, scalar-induced gravitational waves, induced gravitational waves, and so on, in the literature. The produced gravitational waves constitute the stochastic gravitational-wave background. For reviews, see Refs.~\cite{Yuan:2021qgz, Domenech:2021ztg}.

The curvature perturbations around $k\sim10^{14}\ \mathrm{Mpc^{-1}}$ reenter the Hubble horizon in the radiation-dominated era, and the curvature-induced gravitational waves are mostly produced around the horizon reentry. 
We calculate the energy density of gravitational waves produced by the curvature perturbations enhanced around the approximate inflection point to see the detectability of such gravitational waves, following Ref.~\cite{Kohri:2018awv}. 

Before the matter-radiation equality and on subhorizon scales, the energy density fraction for the curvature-induced gravitational waves per $\ln k$ is given by\footnote{
The gravitational-wave energy density truncated at the second order is gauge-dependent~\cite{Hwang:2017oxa, Gong:2019mui, Tomikawa:2019tvi, Lu:2020diy}, but it can be easily upgraded to a gauge-invariant quantity although there are infinite candidates~\cite{Arroja:2009sh, Chang:2020tji}. Ref.~\cite{DeLuca:2019ufz} advocates that the gauge invariant that is equivalent to the naive definition in the synchronous gauge is the physical quantity observable by space-based interferometers like LISA, BBO, and DECIGO. Since the difference between the curvature-induced gravitational waves in the synchronous gauge and the conformal Newtonian gauge vanishes in the late-time limit~\cite{DeLuca:2019ufz, Inomata:2019yww, Yuan:2019fwv}, the emerging consensus is that the energy density of gravitational waves observed in these experiments is well described in the calculations in the conformal Newtonian as well as the synchronous gauges~\cite{Chang:2020iji, Chang:2020mky, Domenech:2020xin} at least for the gravitational waves produced in the radiation-dominated era. See also Refs.~\cite{Cai:2021jbi, Cai:2021ndu}.  We take the conformal Newtonian gauge.}
\begin{align}
    \Omega_{\mathrm{GW}}(\eta, k)=\frac{1}{24}\left(\frac{k}{aH}\right)^2 \overline{\mathcal{P}_h(\eta,k)},
\end{align}
where the bar denotes the oscillation average, and $u = (t+s+1)/2$ and $v=(t-s+1)/2$.
$\mathcal{P}_h$ is the dimensionless power spectrum of gravitational waves, which can be expressed as
\begin{align}
    \overline{\mathcal{P}_h(\eta,k)}=2 \int_0^{\infty} \mathrm{d} t \int_{-1}^1 \mathrm{d}s\left[\frac{t(2+t)\left(s^2-1\right)}{(1-s+t)(1+s+t)}\right]^2 \overline{I^2\left(v, u, k\eta \right)} \mathcal{P}_\mathcal{R}(k v) \mathcal{P}_\mathcal{R}(k u).
\end{align}
Here, we have assumed Gaussian statistics.
The effect of non-Gaussianity of curvature perturbations was studied in Ref.~\cite{Garcia-Bellido:2017aan, Cai:2018dig, Unal:2018yaa,Nakama:2016gzw}.
The probability distribution of curvature perturbations has an exponential tail in the transient USR scenario~\cite{Vennin:2015hra,Pattison:2017mbe,Ezquiaga:2019ftu,Vennin:2020kng,Figueroa:2020jkf,Pattison:2021oen,Figueroa:2021zah,Animali:2022otk,Biagetti:2021eep,Pi:2022ysn}.
The required value of $\mathcal{P}_\mathcal{R}$ for a given amount of PBHs is reduced, so is $\Omega_\text{GW}$.
This is qualitatively similar to the case in Ref.~\cite{Garcia-Bellido:2017aan}.
For the gravitational waves generated in the radiation-dominated era, the integrand $I$ sufficiently long after the horizon reentry is given by~\cite{Espinosa:2018eve, Kohri:2018awv}
\begin{align}
    \overline{I^2\left (t, s, k\eta \rightarrow \infty \right)}=&\left(\frac{1}{k\eta}\right)^{2} \frac{288\left(-5+s^2+t(2+t)\right)^2}{(1-s+t)^6(1+s+t)^6}\Biggl(\frac{\pi^2}{4}\left(-5+s^2+t(2+t)\right)^2 \Theta(t-(\sqrt{3}-1))\nonumber \\
    &+\left(-(t-s+1)(t+s+1)+\frac{1}{2}\left(-5+s^2+t(2+t)\right)\ln \left|\frac{-2+t(2+t)}{3-s^2}\right|\right)^2\Biggr).
\end{align}

The present energy density fraction for the curvature-induced gravitational waves is given by
\begin{align}
    \Omega_\text{GW}(f) = \left( \frac{g_*(T)}{g_*(T_\text{eq})} \right) \left( \frac{g_{*s}(T_\text{eq})}{g_{*s}(T)} \right)^{4/3} \Omega_\text{r} \, \Omega_\text{GW}(\eta_c, 2\pi f),
\end{align}
where $f=k/2\pi$ is the frequency of the gravitational wave, $\Omega_r=\Omega_\text{photon}+\Omega_\text{neutrino}=4.2\times10^{-5}h^{-2}$~\cite{Komatsu_2011} is the present density parameter for radiation, $T$ is to be evaluated at the gravitational-wave production (essentially at the horizon entry of the relevant curvature-perturbation mode), and $\Omega_\text{GW}(\eta_c, k)$ denotes the asymptotic value of $\Omega_\text{GW}(\eta, k)$ well after the production but before the matter-radiation equality. 

\begin{figure}[t]
    \centering
    \includegraphics[width=12cm]{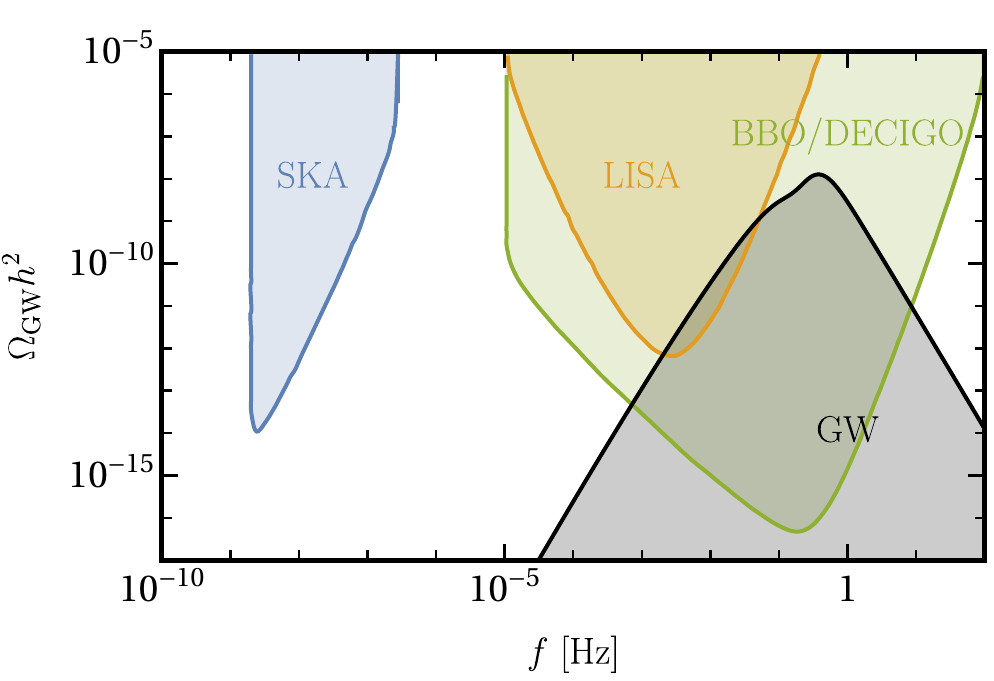}
    \caption{Current energy-density spectrum of the stochastic gravitational wave background induced by the enhanced curvature perturbations (black solid line). The same parameter set~\eqref{eq:parameters} is used as in Figs.~\ref{fg:powerspectrum} and \ref{fg:PBHAbundance}.  The sensitivity curves of detectors such as SKA, LISA, and BBO/DECIGO are also shown by colored lines, which are drawn based on Ref.~\cite{Braglia:2022phb}.}
    \label{fg:PowerSpectrum}
\end{figure}

Fig.~\ref{fg:PowerSpectrum} shows the current energy density of curvature-induced gravitational waves generated by the approximate inflection point in the potential~\eqref{eq:logarithmicpotentialwithinflection} by the black solid line.  It also shows the sensitivity curves of several gravitational-wave detectors: Square Kilometre Array (SKA)~\cite{Zhao:2013bba} as a pulsar timing array (PTA), Laser Interferometer Space Antenna (LISA)~\cite{LISACosmologyWorkingGroup:2022jok}, and Big Bang Observer (BBO)/Deci-Hertz Interferometer Gravitational Wave Observatory (DECIGO)~\cite{Yagi:2011wg}. 
Since the black shaded region for the gravitational waves overlaps with the shaded regions for the sensitivity of LISA (orange) and BBO/DECIGO (green), the curvature-induced gravitational waves are expected to be detected by these planned future gravitational-wave detectors. 
This connection between the 100\% PBH dark matter and LISA was first pointed out in Ref.~\cite{Saito:2008jc} and more recently revisited in Refs.~\cite{Garcia-Bellido:2017aan,Cai:2018dig,Bartolo:2018evs, Bartolo:2018rku, Unal:2018yaa}.

%%%%%%%%%%%%%%%%%%%%%%%%%%%%%%%%%%%%
%%%%%%%%%%%%%%%%%%%%%%%%%%%%%%%%%%%%

%%%%%%%%%%%%%%%%%%%%%%%%%%%%%%%%%%%%%%%%%%%%
\section{Conclusions \label{sec:summary}}
%%%%%%%%%%%%%%%%%%%%%%%%%%%%%%%%%%%%%%%%%%%%

In this paper, we have generalized the polynomial $\alpha$-attractor models of inflation~\cite{Kallosh:2022feu} and studied its predictions on the inflationary observables. 
The characteristic features of the models are the existence of the pole in the kinetic term and the logarithmic terms in the inflaton potential, whose derivative is singular. We have generalized the order of the pole to be arbitrary, including the case of the zeroth order pole, i.e., the absence of the pole. 
We have studied the dependence of $n_s$ and $r$ on the parameters of the model and have found, in particular, that $n_s$ increases further as the order of the pole $p$ goes away from $p=2$. (Remember that one of the original motivations for the polynomial $\alpha$-attractor is to increase $n_s$ compared to the standard $\alpha$-attractor prediction $n_s = 1 - 2/N_e$.) 
Such models with relatively large $n_s$ are still consistent with the latest observations depending on the parameters and the e-folding number. It is interesting that, once we restrict the parameter space so that $n_s$ is not too large, $r$ tends to be of order $10^{-2}$ (see Fig.~\ref{fg:nsr1}), which can be tested in the near future.

We have taken relatively large values of $n_s$ positively in this paper. Nevertheless, it is also interesting to study the possible ways to further extend the models to predict $n_s$ around the central point of the Planck/BICEP/Keck contour.  We have studied two extensions in Sec.~\ref{sec:predictions} and have found that $n_s$ can be indeed around the central value of the constraint (Figs.~\ref{fg:nsr_generalize_order_log} and \ref{fg:nsr_first_order_log}).

Returning to the base model with a relatively large $n_s$, it is advantageous for PBH production scenarios since the deformation of the potential required to enhance the curvature perturbations at small scales decreases the value of $n_s$. 
To demonstrate this point, we have considered the model with the slightly modified logarithmic potential~\eqref{eq:logarithmicpotentialwithinflection} and estimated the PBH abundance and energy density of the curvature-induced gravitational waves.
We have found the parameter set that (i) produces PBHs that explain the whole amount of dark matter, (ii) produces the curvature-induced gravitational waves which can be detected by LISA and BBO/DECIGO, (iii) predicts $n_s$ satisfying the CMB constraints within 2$\sigma$, and (iv) predicts $r$ also satisfying the CMB constraints within 2$\sigma$ but being large, so this model is testable by LiteBIRD, LISA, and BBO/DECIGO in the near future.

Our generalizations of the cosmological attractor models of inflation systematically shift the values of $n_s$ and $r$ to larger values, and they have strong phenomenological implications for the CMB, PBHs and gravitational waves. This requires further theoretical exploration to realize various orders of the pole in the kinetic term and the logarithmic structure in the inflaton potential.  We leave these for future work.

\acknowledgments
T.T.~is grateful to Yuichiro Tada and Junsei Tokuda for useful discussions about the topic of footnote~\ref{fn:quantum_correction}. 
The work of H.M. was supported by JSPS KAKENHI Grant No. JP22KJ1782 and No. JP23K13100.
This work was supported in part by IBS under the project code, IBS-R018-D1~(T.T.).

\bibliographystyle{JHEP}
\bibliography{refs.bib}

\providecommand{\href}[2]{#2}\begingroup\raggedright\begin{thebibliography}{100}

\bibitem{BICEP:2021xfz}
{\scshape BICEP, Keck} collaboration, \emph{{Improved Constraints on Primordial
  Gravitational Waves using Planck, WMAP, and BICEP/Keck Observations through
  the 2018 Observing Season}},
  \href{https://doi.org/10.1103/PhysRevLett.127.151301}{\emph{Phys. Rev. Lett.}
  {\bfseries 127} (2021) 151301}
  [\href{https://arxiv.org/abs/2110.00483}{{\ttfamily 2110.00483}}].

\bibitem{Tristram:2021tvh}
M.~Tristram et~al., \emph{{Improved limits on the tensor-to-scalar ratio using
  BICEP and Planck data}},
  \href{https://doi.org/10.1103/PhysRevD.105.083524}{\emph{Phys. Rev. D}
  {\bfseries 105} (2022) 083524}
  [\href{https://arxiv.org/abs/2112.07961}{{\ttfamily 2112.07961}}].

\bibitem{Planck:2020olo}
{\scshape Planck} collaboration, \emph{{$Planck$ intermediate results. LVII.
  Joint Planck LFI and HFI data processing}},
  \href{https://doi.org/10.1051/0004-6361/202038073}{\emph{Astron. Astrophys.}
  {\bfseries 643} (2020) A42}
  [\href{https://arxiv.org/abs/2007.04997}{{\ttfamily 2007.04997}}].

\bibitem{LiteBIRD:2020khw}
{\scshape LiteBIRD} collaboration, \emph{{LiteBIRD: JAXA's new strategic
  L-class mission for all-sky surveys of cosmic microwave background
  polarization}}, \href{https://doi.org/10.1117/12.2563050}{\emph{Proc. SPIE
  Int. Soc. Opt. Eng.} {\bfseries 11443} (2020) 114432F}
  [\href{https://arxiv.org/abs/2101.12449}{{\ttfamily 2101.12449}}].

\bibitem{Bezrukov:2007ep}
F.L.~Bezrukov and M.~Shaposhnikov, \emph{{The Standard Model Higgs boson as the
  inflaton}}, \href{https://doi.org/10.1016/j.physletb.2007.11.072}{\emph{Phys.
  Lett. B} {\bfseries 659} (2008) 703}
  [\href{https://arxiv.org/abs/0710.3755}{{\ttfamily 0710.3755}}].

\bibitem{Starobinsky:1980te}
A.A.~Starobinsky, \emph{{A New Type of Isotropic Cosmological Models Without
  Singularity}},
  \href{https://doi.org/10.1016/0370-2693(80)90670-X}{\emph{Phys. Lett. B}
  {\bfseries 91} (1980) 99}.

\bibitem{Ellis:2013nxa}
J.~Ellis, D.V.~Nanopoulos and K.A.~Olive, \emph{{Starobinsky-like Inflationary
  Models as Avatars of No-Scale Supergravity}},
  \href{https://doi.org/10.1088/1475-7516/2013/10/009}{\emph{JCAP} {\bfseries
  10} (2013) 009} [\href{https://arxiv.org/abs/1307.3537}{{\ttfamily
  1307.3537}}].

\bibitem{Ferrara:2013rsa}
S.~Ferrara, R.~Kallosh, A.~Linde and M.~Porrati, \emph{{Minimal Supergravity
  Models of Inflation}},
  \href{https://doi.org/10.1103/PhysRevD.88.085038}{\emph{Phys. Rev. D}
  {\bfseries 88} (2013) 085038}
  [\href{https://arxiv.org/abs/1307.7696}{{\ttfamily 1307.7696}}].

\bibitem{Kallosh:2013yoa}
R.~Kallosh, A.~Linde and D.~Roest, \emph{{Superconformal Inflationary
  $\alpha$-Attractors}},
  \href{https://doi.org/10.1007/JHEP11(2013)198}{\emph{JHEP} {\bfseries 11}
  (2013) 198} [\href{https://arxiv.org/abs/1311.0472}{{\ttfamily 1311.0472}}].

\bibitem{Kallosh:2014rga}
R.~Kallosh, A.~Linde and D.~Roest, \emph{{Large field inflation and double
  $\alpha$-attractors}},
  \href{https://doi.org/10.1007/JHEP08(2014)052}{\emph{JHEP} {\bfseries 08}
  (2014) 052} [\href{https://arxiv.org/abs/1405.3646}{{\ttfamily 1405.3646}}].

\bibitem{Carrasco:2015pla}
J.J.M.~Carrasco, R.~Kallosh and A.~Linde, \emph{{$\alpha $-Attractors: Planck,
  LHC and Dark Energy}},
  \href{https://doi.org/10.1007/JHEP10(2015)147}{\emph{JHEP} {\bfseries 10}
  (2015) 147} [\href{https://arxiv.org/abs/1506.01708}{{\ttfamily
  1506.01708}}].

\bibitem{Roest:2015qya}
D.~Roest and M.~Scalisi, \emph{{Cosmological attractors from
  \ensuremath{\alpha}-scale supergravity}},
  \href{https://doi.org/10.1103/PhysRevD.92.043525}{\emph{Phys. Rev. D}
  {\bfseries 92} (2015) 043525}
  [\href{https://arxiv.org/abs/1503.07909}{{\ttfamily 1503.07909}}].

\bibitem{Linde:2015uga}
A.~Linde, \emph{{Single-field $\alpha$-attractors}},
  \href{https://doi.org/10.1088/1475-7516/2015/05/003}{\emph{JCAP} {\bfseries
  05} (2015) 003} [\href{https://arxiv.org/abs/1504.00663}{{\ttfamily
  1504.00663}}].

\bibitem{Scalisi:2015qga}
M.~Scalisi, \emph{{Cosmological $\alpha$-attractors and de Sitter landscape}},
  \href{https://doi.org/10.1007/JHEP12(2015)134}{\emph{JHEP} {\bfseries 12}
  (2015) 134} [\href{https://arxiv.org/abs/1506.01368}{{\ttfamily
  1506.01368}}].

\bibitem{Galante:2014ifa}
M.~Galante, R.~Kallosh, A.~Linde and D.~Roest, \emph{{Unity of Cosmological
  Inflation Attractors}},
  \href{https://doi.org/10.1103/PhysRevLett.114.141302}{\emph{Phys. Rev. Lett.}
  {\bfseries 114} (2015) 141302}
  [\href{https://arxiv.org/abs/1412.3797}{{\ttfamily 1412.3797}}].

\bibitem{Linde:1981mu}
A.D.~Linde, \emph{{A New Inflationary Universe Scenario: A Possible Solution of
  the Horizon, Flatness, Homogeneity, Isotropy and Primordial Monopole
  Problems}}, \href{https://doi.org/10.1016/0370-2693(82)91219-9}{\emph{Phys.
  Lett. B} {\bfseries 108} (1982) 389}.

\bibitem{Albrecht:1982wi}
A.~Albrecht and P.J.~Steinhardt, \emph{{Cosmology for Grand Unified Theories
  with Radiatively Induced Symmetry Breaking}},
  \href{https://doi.org/10.1103/PhysRevLett.48.1220}{\emph{Phys. Rev. Lett.}
  {\bfseries 48} (1982) 1220}.

\bibitem{Boubekeur:2005zm}
L.~Boubekeur and D.H.~Lyth, \emph{{Hilltop inflation}},
  \href{https://doi.org/10.1088/1475-7516/2005/07/010}{\emph{JCAP} {\bfseries
  07} (2005) 010} [\href{https://arxiv.org/abs/hep-ph/0502047}{{\ttfamily
  hep-ph/0502047}}].

\bibitem{Mukhanov:2013tua}
V.~Mukhanov, \emph{{Quantum Cosmological Perturbations: Predictions and
  Observations}},
  \href{https://doi.org/10.1140/epjc/s10052-013-2486-7}{\emph{Eur. Phys. J. C}
  {\bfseries 73} (2013) 2486}
  [\href{https://arxiv.org/abs/1303.3925}{{\ttfamily 1303.3925}}].

\bibitem{Roest:2013fha}
D.~Roest, \emph{{Universality classes of inflation}},
  \href{https://doi.org/10.1088/1475-7516/2014/01/007}{\emph{JCAP} {\bfseries
  01} (2014) 007} [\href{https://arxiv.org/abs/1309.1285}{{\ttfamily
  1309.1285}}].

\bibitem{Garcia-Bellido:2014gna}
J.~Garcia-Bellido and D.~Roest, \emph{{Large-$N$ running of the spectral index
  of inflation}}, \href{https://doi.org/10.1103/PhysRevD.89.103527}{\emph{Phys.
  Rev. D} {\bfseries 89} (2014) 103527}
  [\href{https://arxiv.org/abs/1402.2059}{{\ttfamily 1402.2059}}].

\bibitem{Binetruy:2014zya}
P.~Binetruy, E.~Kiritsis, J.~Mabillard, M.~Pieroni and C.~Rosset,
  \emph{{Universality classes for models of inflation}},
  \href{https://doi.org/10.1088/1475-7516/2015/04/033}{\emph{JCAP} {\bfseries
  04} (2015) 033} [\href{https://arxiv.org/abs/1407.0820}{{\ttfamily
  1407.0820}}].

\bibitem{Broy:2015qna}
B.J.~Broy, M.~Galante, D.~Roest and A.~Westphal, \emph{{Pole inflation
  \textemdash{} Shift symmetry and universal corrections}},
  \href{https://doi.org/10.1007/JHEP12(2015)149}{\emph{JHEP} {\bfseries 12}
  (2015) 149} [\href{https://arxiv.org/abs/1507.02277}{{\ttfamily
  1507.02277}}].

\bibitem{Terada:2016nqg}
T.~Terada, \emph{{Generalized Pole Inflation: Hilltop, Natural, and Chaotic
  Inflationary Attractors}},
  \href{https://doi.org/10.1016/j.physletb.2016.07.058}{\emph{Phys. Lett. B}
  {\bfseries 760} (2016) 674}
  [\href{https://arxiv.org/abs/1602.07867}{{\ttfamily 1602.07867}}].

\bibitem{Kallosh:2022feu}
R.~Kallosh and A.~Linde, \emph{{Polynomial \ensuremath{\alpha}-attractors}},
  \href{https://doi.org/10.1088/1475-7516/2022/04/017}{\emph{JCAP} {\bfseries
  04} (2022) 017} [\href{https://arxiv.org/abs/2202.06492}{{\ttfamily
  2202.06492}}].

\bibitem{Linde:1983gd}
A.D.~Linde, \emph{{Chaotic Inflation}},
  \href{https://doi.org/10.1016/0370-2693(83)90837-7}{\emph{Phys. Lett. B}
  {\bfseries 129} (1983) 177}.

\bibitem{Bhattacharya:2022akq}
S.~Bhattacharya, K.~Dutta, M.R.~Gangopadhyay and A.~Maharana,
  \emph{{$\alpha$-attractor inflation: Models and Predictions}},
  \href{https://arxiv.org/abs/2212.13363}{{\ttfamily 2212.13363}}.

\bibitem{ACT:2020gnv}
{\scshape ACT} collaboration, \emph{{The Atacama Cosmology Telescope: DR4 Maps
  and Cosmological Parameters}},
  \href{https://doi.org/10.1088/1475-7516/2020/12/047}{\emph{JCAP} {\bfseries
  12} (2020) 047} [\href{https://arxiv.org/abs/2007.07288}{{\ttfamily
  2007.07288}}].

\bibitem{Braglia:2020bym}
M.~Braglia, W.T.~Emond, F.~Finelli, A.E.~Gumrukcuoglu and K.~Koyama,
  \emph{{Unified framework for early dark energy from $\alpha$-attractors}},
  \href{https://doi.org/10.1103/PhysRevD.102.083513}{\emph{Phys. Rev. D}
  {\bfseries 102} (2020) 083513}
  [\href{https://arxiv.org/abs/2005.14053}{{\ttfamily 2005.14053}}].

\bibitem{Jiang:2022uyg}
J.-Q.~Jiang and Y.-S.~Piao, \emph{{Toward early dark energy and ns=1 with
  Planck, ACT, and SPT observations}},
  \href{https://doi.org/10.1103/PhysRevD.105.103514}{\emph{Phys. Rev. D}
  {\bfseries 105} (2022) 103514}
  [\href{https://arxiv.org/abs/2202.13379}{{\ttfamily 2202.13379}}].

\bibitem{Cruz:2022oqk}
J.S.~Cruz, F.~Niedermann and M.S.~Sloth, \emph{{A grounded perspective on New
  Early Dark Energy using ACT, SPT, and BICEP/Keck}},
  \href{https://arxiv.org/abs/2209.02708}{{\ttfamily 2209.02708}}.

\bibitem{Silk:1986vc}
J.~Silk and M.S.~Turner, \emph{{Double Inflation}},
  \href{https://doi.org/10.1103/PhysRevD.35.419}{\emph{Phys. Rev. D} {\bfseries
  35} (1987) 419}.

\bibitem{Hawking:1971ei}
S.~Hawking, \emph{{Gravitationally collapsed objects of very low mass}},
  {\emph{Mon. Not. Roy. Astron. Soc.} {\bfseries 152} (1971) 75}.

\bibitem{Carr:1974nx}
B.J.~Carr and S.W.~Hawking, \emph{{Black holes in the early Universe}},
  \href{https://doi.org/10.1093/mnras/168.2.399}{\emph{Mon. Not. Roy. Astron.
  Soc.} {\bfseries 168} (1974) 399}.

\bibitem{Inoue:2001zt}
S.~Inoue and J.~Yokoyama, \emph{{Curvature perturbation at the local extremum
  of the inflaton's potential}},
  \href{https://doi.org/10.1016/S0370-2693(01)01369-7}{\emph{Phys. Lett. B}
  {\bfseries 524} (2002) 15}
  [\href{https://arxiv.org/abs/hep-ph/0104083}{{\ttfamily hep-ph/0104083}}].

\bibitem{Tsamis:2003px}
N.C.~Tsamis and R.P.~Woodard, \emph{{Improved estimates of cosmological
  perturbations}},
  \href{https://doi.org/10.1103/PhysRevD.69.084005}{\emph{Phys. Rev. D}
  {\bfseries 69} (2004) 084005}
  [\href{https://arxiv.org/abs/astro-ph/0307463}{{\ttfamily
  astro-ph/0307463}}].

\bibitem{Kinney:2005vj}
W.H.~Kinney, \emph{{Horizon crossing and inflation with large eta}},
  \href{https://doi.org/10.1103/PhysRevD.72.023515}{\emph{Phys. Rev. D}
  {\bfseries 72} (2005) 023515}
  [\href{https://arxiv.org/abs/gr-qc/0503017}{{\ttfamily gr-qc/0503017}}].

\bibitem{Ezquiaga:2017fvi}
J.M.~Ezquiaga, J.~Garcia-Bellido and E.~Ruiz~Morales, \emph{{Primordial Black
  Hole production in Critical Higgs Inflation}},
  \href{https://doi.org/10.1016/j.physletb.2017.11.039}{\emph{Phys. Lett. B}
  {\bfseries 776} (2018) 345}
  [\href{https://arxiv.org/abs/1705.04861}{{\ttfamily 1705.04861}}].

\bibitem{Motohashi:2017kbs}
H.~Motohashi and W.~Hu, \emph{{Primordial Black Holes and Slow-Roll
  Violation}}, \href{https://doi.org/10.1103/PhysRevD.96.063503}{\emph{Phys.
  Rev. D} {\bfseries 96} (2017) 063503}
  [\href{https://arxiv.org/abs/1706.06784}{{\ttfamily 1706.06784}}].

\bibitem{Germani:2017bcs}
C.~Germani and T.~Prokopec, \emph{{On primordial black holes from an inflection
  point}}, \href{https://doi.org/10.1016/j.dark.2017.09.001}{\emph{Phys. Dark
  Univ.} {\bfseries 18} (2017) 6}
  [\href{https://arxiv.org/abs/1706.04226}{{\ttfamily 1706.04226}}].

\bibitem{Iacconi:2021ltm}
L.~Iacconi, H.~Assadullahi, M.~Fasiello and D.~Wands, \emph{{Revisiting
  small-scale fluctuations in \ensuremath{\alpha}-attractor models of
  inflation}}, \href{https://doi.org/10.1088/1475-7516/2022/06/007}{\emph{JCAP}
  {\bfseries 06} (2022) 007}
  [\href{https://arxiv.org/abs/2112.05092}{{\ttfamily 2112.05092}}].

\bibitem{Kallosh:2022ggf}
R.~Kallosh and A.~Linde, \emph{{Hybrid cosmological attractors}},
  \href{https://doi.org/10.1103/PhysRevD.106.023522}{\emph{Phys. Rev. D}
  {\bfseries 106} (2022) 023522}
  [\href{https://arxiv.org/abs/2204.02425}{{\ttfamily 2204.02425}}].

\bibitem{Braglia:2022phb}
M.~Braglia, A.~Linde, R.~Kallosh and F.~Finelli, \emph{{Hybrid
  $\alpha$-attractors, primordial black holes and gravitational wave
  backgrounds}},  \href{https://arxiv.org/abs/2211.14262}{{\ttfamily
  2211.14262}}.

\bibitem{Planck:2018jri}
{\scshape Planck} collaboration, \emph{{Planck 2018 results. X. Constraints on
  inflation}}, \href{https://doi.org/10.1051/0004-6361/201833887}{\emph{Astron.
  Astrophys.} {\bfseries 641} (2020) A10}
  [\href{https://arxiv.org/abs/1807.06211}{{\ttfamily 1807.06211}}].

\bibitem{bicepkeck}
\url{http://bicepkeck.org/bk18_2021_release.html}.

\bibitem{Aldabergenov:2023yrk}
Y.~Aldabergenov and S.V.~Ketov, \emph{{Primordial black holes from
  Volkov-Akulov-Starobinsky supergravity}},
  \href{https://arxiv.org/abs/2301.12750}{{\ttfamily 2301.12750}}.

\bibitem{Wu:2021zta}
L.~Wu, Y.~Gong and T.~Li, \emph{{Primordial black holes and secondary
  gravitational waves from string inspired general no-scale supergravity}},
  \href{https://doi.org/10.1103/PhysRevD.104.123544}{\emph{Phys. Rev. D}
  {\bfseries 104} (2021) 123544}
  [\href{https://arxiv.org/abs/2105.07694}{{\ttfamily 2105.07694}}].

\bibitem{Kristiano:2022maq}
J.~Kristiano and J.~Yokoyama, \emph{{Ruling Out Primordial Black Hole Formation
  From Single-Field Inflation}},
  \href{https://arxiv.org/abs/2211.03395}{{\ttfamily 2211.03395}}.

\bibitem{Inomata:2022yte}
K.~Inomata, M.~Braglia and X.~Chen, \emph{{Questions on calculation of
  primordial power spectrum with large spikes: the resonance model case}},
  \href{https://arxiv.org/abs/2211.02586}{{\ttfamily 2211.02586}}.

\bibitem{Cheng:2021lif}
S.-L.~Cheng, D.-S.~Lee and K.-W.~Ng, \emph{{Power spectrum of primordial
  perturbations during ultra-slow-roll inflation with back reaction effects}},
  \href{https://doi.org/10.1016/j.physletb.2022.136956}{\emph{Phys. Lett. B}
  {\bfseries 827} (2022) 136956}
  [\href{https://arxiv.org/abs/2106.09275}{{\ttfamily 2106.09275}}].

\bibitem{Riotto:2023hoz}
A.~Riotto, \emph{{The Primordial Black Hole Formation from Single-Field
  Inflation is Not Ruled Out}},
  \href{https://arxiv.org/abs/2301.00599}{{\ttfamily 2301.00599}}.

\bibitem{Stewart:1993bc}
E.D.~Stewart and D.H.~Lyth, \emph{{A More accurate analytic calculation of the
  spectrum of cosmological perturbations produced during inflation}},
  \href{https://doi.org/10.1016/0370-2693(93)90379-V}{\emph{Phys. Lett. B}
  {\bfseries 302} (1993) 171}
  [\href{https://arxiv.org/abs/gr-qc/9302019}{{\ttfamily gr-qc/9302019}}].

\bibitem{Mukhanov:1988jd}
V.F.~Mukhanov, \emph{{Quantum Theory of Gauge Invariant Cosmological
  Perturbations}}, {\emph{Sov. Phys. JETP} {\bfseries 67} (1988) 1297}.

\bibitem{Sasaki:1986hm}
M.~Sasaki, \emph{{Large Scale Quantum Fluctuations in the Inflationary
  Universe}}, \href{https://doi.org/10.1143/PTP.76.1036}{\emph{Prog. Theor.
  Phys.} {\bfseries 76} (1986) 1036}.

\bibitem{Ballesteros:2017fsr}
G.~Ballesteros and M.~Taoso, \emph{{Primordial black hole dark matter from
  single field inflation}},
  \href{https://doi.org/10.1103/PhysRevD.97.023501}{\emph{Phys. Rev. D}
  {\bfseries 97} (2018) 023501}
  [\href{https://arxiv.org/abs/1709.05565}{{\ttfamily 1709.05565}}].

\bibitem{Drees:2019xpp}
M.~Drees and Y.~Xu, \emph{{Overshooting, Critical Higgs Inflation and Second
  Order Gravitational Wave Signatures}},
  \href{https://doi.org/10.1140/epjc/s10052-021-08976-2}{\emph{Eur. Phys. J. C}
  {\bfseries 81} (2021) 182}
  [\href{https://arxiv.org/abs/1905.13581}{{\ttfamily 1905.13581}}].

\bibitem{Byrnes:2018txb}
C.T.~Byrnes, P.S.~Cole and S.P.~Patil, \emph{{Steepest growth of the power
  spectrum and primordial black holes}},
  \href{https://doi.org/10.1088/1475-7516/2019/06/028}{\emph{JCAP} {\bfseries
  06} (2019) 028} [\href{https://arxiv.org/abs/1811.11158}{{\ttfamily
  1811.11158}}].

\bibitem{Carrilho:2019oqg}
P.~Carrilho, K.A.~Malik and D.J.~Mulryne, \emph{{Dissecting the growth of the
  power spectrum for primordial black holes}},
  \href{https://doi.org/10.1103/PhysRevD.100.103529}{\emph{Phys. Rev. D}
  {\bfseries 100} (2019) 103529}
  [\href{https://arxiv.org/abs/1907.05237}{{\ttfamily 1907.05237}}].

\bibitem{Cole:2022xqc}
P.S.~Cole, A.D.~Gow, C.T.~Byrnes and S.P.~Patil, \emph{{Steepest growth
  re-examined: repercussions for primordial black hole formation}},
  \href{https://arxiv.org/abs/2204.07573}{{\ttfamily 2204.07573}}.

\bibitem{Liu:2020oqe}
J.~Liu, Z.-K.~Guo and R.-G.~Cai, \emph{{Analytical approximation of the scalar
  spectrum in the ultraslow-roll inflationary models}},
  \href{https://doi.org/10.1103/PhysRevD.101.083535}{\emph{Phys. Rev. D}
  {\bfseries 101} (2020) 083535}
  [\href{https://arxiv.org/abs/2003.02075}{{\ttfamily 2003.02075}}].

\bibitem{Carr:1975qj}
B.J.~Carr, \emph{{The Primordial black hole mass spectrum}},
  \href{https://doi.org/10.1086/153853}{\emph{Astrophys. J.} {\bfseries 201}
  (1975) 1}.

\bibitem{1986ApJ...304...15B}
J.M.~{Bardeen}, J.R.~{Bond}, N.~{Kaiser} and A.S.~{Szalay}, \emph{{The
  statistics of peaks of Gaussian random fields}},
  \href{https://doi.org/10.1086/164143}{\emph{Astrophys. J.} {\bfseries 304}
  (1986) 15}.

\bibitem{Yoo:2018kvb}
C.-M.~Yoo, T.~Harada, J.~Garriga and K.~Kohri, \emph{{Primordial black hole
  abundance from random Gaussian curvature perturbations and a local density
  threshold}}, \href{https://doi.org/10.1093/ptep/pty120}{\emph{PTEP}
  {\bfseries 2018} (2018) 123E01}
  [\href{https://arxiv.org/abs/1805.03946}{{\ttfamily 1805.03946}}].

\bibitem{Germani:2018jgr}
C.~Germani and I.~Musco, \emph{{Abundance of Primordial Black Holes Depends on
  the Shape of the Inflationary Power Spectrum}},
  \href{https://doi.org/10.1103/PhysRevLett.122.141302}{\emph{Phys. Rev. Lett.}
  {\bfseries 122} (2019) 141302}
  [\href{https://arxiv.org/abs/1805.04087}{{\ttfamily 1805.04087}}].

\bibitem{Yoo:2020dkz}
C.-M.~Yoo, T.~Harada, S.~Hirano and K.~Kohri, \emph{{Abundance of Primordial
  Black Holes in Peak Theory for an Arbitrary Power Spectrum}},
  \href{https://doi.org/10.1093/ptep/ptaa155}{\emph{PTEP} {\bfseries 2021}
  (2021) 013E02} [\href{https://arxiv.org/abs/2008.02425}{{\ttfamily
  2008.02425}}].

\bibitem{Niemeyer:1997mt}
J.C.~Niemeyer and K.~Jedamzik, \emph{{Near-critical gravitational collapse and
  the initial mass function of primordial black holes}},
  \href{https://doi.org/10.1103/PhysRevLett.80.5481}{\emph{Phys. Rev. Lett.}
  {\bfseries 80} (1998) 5481}
  [\href{https://arxiv.org/abs/astro-ph/9709072}{{\ttfamily
  astro-ph/9709072}}].

\bibitem{Yokoyama:1998xd}
J.~Yokoyama, \emph{{Cosmological constraints on primordial black holes produced
  in the near critical gravitational collapse}},
  \href{https://doi.org/10.1103/PhysRevD.58.107502}{\emph{Phys. Rev. D}
  {\bfseries 58} (1998) 107502}
  [\href{https://arxiv.org/abs/gr-qc/9804041}{{\ttfamily gr-qc/9804041}}].

\bibitem{Green:1999xm}
A.M.~Green and A.R.~Liddle, \emph{{Critical collapse and the primordial black
  hole initial mass function}},
  \href{https://doi.org/10.1103/PhysRevD.60.063509}{\emph{Phys. Rev. D}
  {\bfseries 60} (1999) 063509}
  [\href{https://arxiv.org/abs/astro-ph/9901268}{{\ttfamily
  astro-ph/9901268}}].

\bibitem{Kuhnel:2015vtw}
F.~K\"uhnel, C.~Rampf and M.~Sandstad, \emph{{Effects of Critical Collapse on
  Primordial Black-Hole Mass Spectra}},
  \href{https://doi.org/10.1140/epjc/s10052-016-3945-8}{\emph{Eur. Phys. J. C}
  {\bfseries 76} (2016) 93} [\href{https://arxiv.org/abs/1512.00488}{{\ttfamily
  1512.00488}}].

\bibitem{DeLuca:2019qsy}
V.~De~Luca, G.~Franciolini, A.~Kehagias, M.~Peloso, A.~Riotto and C.~\"Unal,
  \emph{{The Ineludible non-Gaussianity of the Primordial Black Hole
  Abundance}}, \href{https://doi.org/10.1088/1475-7516/2019/07/048}{\emph{JCAP}
  {\bfseries 07} (2019) 048}
  [\href{https://arxiv.org/abs/1904.00970}{{\ttfamily 1904.00970}}].

\bibitem{Atal:2019cdz}
V.~Atal, J.~Garriga and A.~Marcos-Caballero, \emph{{Primordial black hole
  formation with non-Gaussian curvature perturbations}},
  \href{https://doi.org/10.1088/1475-7516/2019/09/073}{\emph{JCAP} {\bfseries
  09} (2019) 073} [\href{https://arxiv.org/abs/1905.13202}{{\ttfamily
  1905.13202}}].

\bibitem{Biagetti:2021eep}
M.~Biagetti, V.~De~Luca, G.~Franciolini, A.~Kehagias and A.~Riotto, \emph{{The
  formation probability of primordial black holes}},
  \href{https://doi.org/10.1016/j.physletb.2021.136602}{\emph{Phys. Lett. B}
  {\bfseries 820} (2021) 136602}
  [\href{https://arxiv.org/abs/2105.07810}{{\ttfamily 2105.07810}}].

\bibitem{Kitajima:2021fpq}
N.~Kitajima, Y.~Tada, S.~Yokoyama and C.-M.~Yoo, \emph{{Primordial black holes
  in peak theory with a non-Gaussian tail}},
  \href{https://doi.org/10.1088/1475-7516/2021/10/053}{\emph{JCAP} {\bfseries
  10} (2021) 053} [\href{https://arxiv.org/abs/2109.00791}{{\ttfamily
  2109.00791}}].

\bibitem{Escriva:2022pnz}
A.~Escriv\`a, Y.~Tada, S.~Yokoyama and C.-M.~Yoo, \emph{{Simulation of
  primordial black holes with large negative non-Gaussianity}},
  \href{https://doi.org/10.1088/1475-7516/2022/05/012}{\emph{JCAP} {\bfseries
  05} (2022) 012} [\href{https://arxiv.org/abs/2202.01028}{{\ttfamily
  2202.01028}}].

\bibitem{Pi:2022ysn}
S.~Pi and M.~Sasaki, \emph{{Logarithmic Duality of the Curvature
  Perturbation}},  \href{https://arxiv.org/abs/2211.13932}{{\ttfamily
  2211.13932}}.

\bibitem{Khlopov:2008qy}
M.Y.~Khlopov, \emph{{Primordial Black Holes}},
  \href{https://doi.org/10.1088/1674-4527/10/6/001}{\emph{Res. Astron.
  Astrophys.} {\bfseries 10} (2010) 495}
  [\href{https://arxiv.org/abs/0801.0116}{{\ttfamily 0801.0116}}].

\bibitem{Sasaki:2018dmp}
M.~Sasaki, T.~Suyama, T.~Tanaka and S.~Yokoyama, \emph{{Primordial black
  holes\textemdash{}perspectives in gravitational wave astronomy}},
  \href{https://doi.org/10.1088/1361-6382/aaa7b4}{\emph{Class. Quant. Grav.}
  {\bfseries 35} (2018) 063001}
  [\href{https://arxiv.org/abs/1801.05235}{{\ttfamily 1801.05235}}].

\bibitem{Carr:2020gox}
B.~Carr, K.~Kohri, Y.~Sendouda and J.~Yokoyama, \emph{{Constraints on
  primordial black holes}},
  \href{https://doi.org/10.1088/1361-6633/ac1e31}{\emph{Rept. Prog. Phys.}
  {\bfseries 84} (2021) 116902}
  [\href{https://arxiv.org/abs/2002.12778}{{\ttfamily 2002.12778}}].

\bibitem{Carr:2020xqk}
B.~Carr and F.~Kuhnel, \emph{{Primordial Black Holes as Dark Matter: Recent
  Developments}},
  \href{https://doi.org/10.1146/annurev-nucl-050520-125911}{\emph{Ann. Rev.
  Nucl. Part. Sci.} {\bfseries 70} (2020) 355}
  [\href{https://arxiv.org/abs/2006.02838}{{\ttfamily 2006.02838}}].

\bibitem{Escriva:2022duf}
A.~Escriv\`a, F.~Kuhnel and Y.~Tada, \emph{{Primordial Black Holes}},
  \href{https://arxiv.org/abs/2211.05767}{{\ttfamily 2211.05767}}.

\bibitem{Inomata:2017okj}
K.~Inomata, M.~Kawasaki, K.~Mukaida, Y.~Tada and T.T.~Yanagida,
  \emph{{Inflationary Primordial Black Holes as All Dark Matter}},
  \href{https://doi.org/10.1103/PhysRevD.96.043504}{\emph{Phys. Rev. D}
  {\bfseries 96} (2017) 043504}
  [\href{https://arxiv.org/abs/1701.02544}{{\ttfamily 1701.02544}}].

\bibitem{Musco:2004ak}
I.~Musco, J.C.~Miller and L.~Rezzolla, \emph{{Computations of primordial black
  hole formation}},
  \href{https://doi.org/10.1088/0264-9381/22/7/013}{\emph{Class. Quant. Grav.}
  {\bfseries 22} (2005) 1405}
  [\href{https://arxiv.org/abs/gr-qc/0412063}{{\ttfamily gr-qc/0412063}}].

\bibitem{Polnarev:2006aa}
A.G.~Polnarev and I.~Musco, \emph{{Curvature profiles as initial conditions for
  primordial black hole formation}},
  \href{https://doi.org/10.1088/0264-9381/24/6/003}{\emph{Class. Quant. Grav.}
  {\bfseries 24} (2007) 1405}
  [\href{https://arxiv.org/abs/gr-qc/0605122}{{\ttfamily gr-qc/0605122}}].

\bibitem{Musco:2008hv}
I.~Musco, J.C.~Miller and A.G.~Polnarev, \emph{{Primordial black hole formation
  in the radiative era: Investigation of the critical nature of the collapse}},
  \href{https://doi.org/10.1088/0264-9381/26/23/235001}{\emph{Class. Quant.
  Grav.} {\bfseries 26} (2009) 235001}
  [\href{https://arxiv.org/abs/0811.1452}{{\ttfamily 0811.1452}}].

\bibitem{Musco:2012au}
I.~Musco and J.C.~Miller, \emph{{Primordial black hole formation in the early
  universe: critical behaviour and self-similarity}},
  \href{https://doi.org/10.1088/0264-9381/30/14/145009}{\emph{Class. Quant.
  Grav.} {\bfseries 30} (2013) 145009}
  [\href{https://arxiv.org/abs/1201.2379}{{\ttfamily 1201.2379}}].

\bibitem{Harada:2013epa}
T.~Harada, C.-M.~Yoo and K.~Kohri, \emph{{Threshold of primordial black hole
  formation}}, \href{https://doi.org/10.1103/PhysRevD.88.084051}{\emph{Phys.
  Rev. D} {\bfseries 88} (2013) 084051}
  [\href{https://arxiv.org/abs/1309.4201}{{\ttfamily 1309.4201}}].

\bibitem{Kohri:2018qtx}
K.~Kohri and T.~Terada, \emph{{Primordial Black Hole Dark Matter and LIGO/Virgo
  Merger Rate from Inflation with Running Spectral Indices: Formation in the
  Matter- and/or Radiation-Dominated Universe}},
  \href{https://doi.org/10.1088/1361-6382/aaea18}{\emph{Class. Quant. Grav.}
  {\bfseries 35} (2018) 235017}
  [\href{https://arxiv.org/abs/1802.06785}{{\ttfamily 1802.06785}}].

\bibitem{Graham:2015apa}
P.W.~Graham, S.~Rajendran and J.~Varela, \emph{{Dark Matter Triggers of
  Supernovae}}, \href{https://doi.org/10.1103/PhysRevD.92.063007}{\emph{Phys.
  Rev. D} {\bfseries 92} (2015) 063007}
  [\href{https://arxiv.org/abs/1505.04444}{{\ttfamily 1505.04444}}].

\bibitem{Capela:2013yf}
F.~Capela, M.~Pshirkov and P.~Tinyakov, \emph{{Constraints on primordial black
  holes as dark matter candidates from capture by neutron stars}},
  \href{https://doi.org/10.1103/PhysRevD.87.123524}{\emph{Phys. Rev. D}
  {\bfseries 87} (2013) 123524}
  [\href{https://arxiv.org/abs/1301.4984}{{\ttfamily 1301.4984}}].

\bibitem{Montero-Camacho:2019jte}
P.~Montero-Camacho, X.~Fang, G.~Vasquez, M.~Silva and C.M.~Hirata,
  \emph{{Revisiting constraints on asteroid-mass primordial black holes as dark
  matter candidates}},
  \href{https://doi.org/10.1088/1475-7516/2019/08/031}{\emph{JCAP} {\bfseries
  08} (2019) 031} [\href{https://arxiv.org/abs/1906.05950}{{\ttfamily
  1906.05950}}].

\bibitem{Niikura:2017zjd}
H.~Niikura et~al., \emph{{Microlensing constraints on primordial black holes
  with Subaru/HSC Andromeda observations}},
  \href{https://doi.org/10.1038/s41550-019-0723-1}{\emph{Nature Astron.}
  {\bfseries 3} (2019) 524} [\href{https://arxiv.org/abs/1701.02151}{{\ttfamily
  1701.02151}}].

\bibitem{Katz:2018zrn}
A.~Katz, J.~Kopp, S.~Sibiryakov and W.~Xue, \emph{{Femtolensing by Dark Matter
  Revisited}}, \href{https://doi.org/10.1088/1475-7516/2018/12/005}{\emph{JCAP}
  {\bfseries 12} (2018) 005}
  [\href{https://arxiv.org/abs/1807.11495}{{\ttfamily 1807.11495}}].

\bibitem{Inomata:2018cht}
K.~Inomata, M.~Kawasaki, K.~Mukaida and T.T.~Yanagida, \emph{{Double inflation
  as a single origin of primordial black holes for all dark matter and LIGO
  observations}}, \href{https://doi.org/10.1103/PhysRevD.97.043514}{\emph{Phys.
  Rev. D} {\bfseries 97} (2018) 043514}
  [\href{https://arxiv.org/abs/1711.06129}{{\ttfamily 1711.06129}}].

\bibitem{gould1992femtolensing}
A.~Gould, \emph{{Femtolensing of gamma-ray bursters}}, {\emph{The Astrophysical
  Journal} {\bfseries 386} (1992) L5}.

\bibitem{ReyIdler:2022unr}
J.L.~Rey~Idler, \emph{{Primordial black holes from inflation and their
  gravitational wave signals}}, Ph.D. thesis, U. Autonoma, Madrid (main), 2022.

\bibitem{Bartolo:2018rku}
N.~Bartolo, V.~De~Luca, G.~Franciolini, M.~Peloso, D.~Racco and A.~Riotto,
  \emph{{Testing primordial black holes as dark matter with LISA}},
  \href{https://doi.org/10.1103/PhysRevD.99.103521}{\emph{Phys. Rev. D}
  {\bfseries 99} (2019) 103521}
  [\href{https://arxiv.org/abs/1810.12224}{{\ttfamily 1810.12224}}].

\bibitem{Villanueva-Domingo:2021spv}
P.~Villanueva-Domingo, O.~Mena and S.~Palomares-Ruiz, \emph{{A brief review on
  primordial black holes as dark matter}},
  \href{https://doi.org/10.3389/fspas.2021.681084}{\emph{Front. Astron. Space
  Sci.} {\bfseries 8} (2021) 87}
  [\href{https://arxiv.org/abs/2103.12087}{{\ttfamily 2103.12087}}].

\bibitem{Carr:2009jm}
B.J.~Carr, K.~Kohri, Y.~Sendouda and J.~Yokoyama, \emph{{New cosmological
  constraints on primordial black holes}},
  \href{https://doi.org/10.1103/PhysRevD.81.104019}{\emph{Phys. Rev. D}
  {\bfseries 81} (2010) 104019}
  [\href{https://arxiv.org/abs/0912.5297}{{\ttfamily 0912.5297}}].

\bibitem{Clark:2016nst}
S.~Clark, B.~Dutta, Y.~Gao, L.E.~Strigari and S.~Watson, \emph{{Planck
  Constraint on Relic Primordial Black Holes}},
  \href{https://doi.org/10.1103/PhysRevD.95.083006}{\emph{Phys. Rev. D}
  {\bfseries 95} (2017) 083006}
  [\href{https://arxiv.org/abs/1612.07738}{{\ttfamily 1612.07738}}].

\bibitem{Macho:2000nvd}
{\scshape Macho} collaboration, \emph{{MACHO project limits on black hole dark
  matter in the 1-30 solar mass range}},
  \href{https://doi.org/10.1086/319636}{\emph{Astrophys. J. Lett.} {\bfseries
  550} (2001) L169} [\href{https://arxiv.org/abs/astro-ph/0011506}{{\ttfamily
  astro-ph/0011506}}].

\bibitem{EROS-2:2006ryy}
{\scshape EROS-2} collaboration, \emph{{Limits on the Macho Content of the
  Galactic Halo from the EROS-2 Survey of the Magellanic Clouds}},
  \href{https://doi.org/10.1051/0004-6361:20066017}{\emph{Astron. Astrophys.}
  {\bfseries 469} (2007) 387}
  [\href{https://arxiv.org/abs/astro-ph/0607207}{{\ttfamily
  astro-ph/0607207}}].

\bibitem{Griest:2013aaa}
K.~Griest, A.M.~Cieplak and M.J.~Lehner, \emph{{Experimental Limits on
  Primordial Black Hole Dark Matter from the First 2 yr of Kepler Data}},
  \href{https://doi.org/10.1088/0004-637X/786/2/158}{\emph{Astrophys. J.}
  {\bfseries 786} (2014) 158}
  [\href{https://arxiv.org/abs/1307.5798}{{\ttfamily 1307.5798}}].

\bibitem{Oguri:2017ock}
M.~Oguri, J.M.~Diego, N.~Kaiser, P.L.~Kelly and T.~Broadhurst,
  \emph{{Understanding caustic crossings in giant arcs: characteristic scales,
  event rates, and constraints on compact dark matter}},
  \href{https://doi.org/10.1103/PhysRevD.97.023518}{\emph{Phys. Rev. D}
  {\bfseries 97} (2018) 023518}
  [\href{https://arxiv.org/abs/1710.00148}{{\ttfamily 1710.00148}}].

\bibitem{Niikura:2019kqi}
H.~Niikura, M.~Takada, S.~Yokoyama, T.~Sumi and S.~Masaki, \emph{{Constraints
  on Earth-mass primordial black holes from OGLE 5-year microlensing events}},
  \href{https://doi.org/10.1103/PhysRevD.99.083503}{\emph{Phys. Rev. D}
  {\bfseries 99} (2019) 083503}
  [\href{https://arxiv.org/abs/1901.07120}{{\ttfamily 1901.07120}}].

\bibitem{Croon:2020ouk}
D.~Croon, D.~McKeen, N.~Raj and Z.~Wang, \emph{{Subaru-HSC through a different
  lens: Microlensing by extended dark matter structures}},
  \href{https://doi.org/10.1103/PhysRevD.102.083021}{\emph{Phys. Rev. D}
  {\bfseries 102} (2020) 083021}
  [\href{https://arxiv.org/abs/2007.12697}{{\ttfamily 2007.12697}}].

\bibitem{Serpico:2020ehh}
P.D.~Serpico, V.~Poulin, D.~Inman and K.~Kohri, \emph{{Cosmic microwave
  background bounds on primordial black holes including dark matter halo
  accretion}},
  \href{https://doi.org/10.1103/PhysRevResearch.2.023204}{\emph{Phys. Rev.
  Res.} {\bfseries 2} (2020) 023204}
  [\href{https://arxiv.org/abs/2002.10771}{{\ttfamily 2002.10771}}].

\bibitem{MonroyRodriguez}
M.A.~Monroy-Rodríguez and C.~Allen, \emph{{The end of the MACHO era-
  revisited: new limits on MACHO masses from halo wide binaries}},
  \href{https://doi.org/10.1088/0004-637X/790/2/159}{\emph{American
  Astronomical Society} {\bfseries 790} (2014) 159}
  [\href{https://arxiv.org/abs/11406.5169}{{\ttfamily 11406.5169}}].

\bibitem{Brandt:2016aco}
T.D.~Brandt, \emph{{Constraints on MACHO Dark Matter from Compact Stellar
  Systems in Ultra-Faint Dwarf Galaxies}},
  \href{https://doi.org/10.3847/2041-8205/824/2/L31}{\emph{Astrophys. J. Lett.}
  {\bfseries 824} (2016) L31}
  [\href{https://arxiv.org/abs/1605.03665}{{\ttfamily 1605.03665}}].

\bibitem{Murgia:2019duy}
R.~Murgia, G.~Scelfo, M.~Viel and A.~Raccanelli,
  \emph{{Lyman-\ensuremath{\alpha} Forest Constraints on Primordial Black Holes
  as Dark Matter}},
  \href{https://doi.org/10.1103/PhysRevLett.123.071102}{\emph{Phys. Rev. Lett.}
  {\bfseries 123} (2019) 071102}
  [\href{https://arxiv.org/abs/1903.10509}{{\ttfamily 1903.10509}}].

\bibitem{Kavanagh:2018ggo}
B.J.~Kavanagh, D.~Gaggero and G.~Bertone, \emph{{Merger rate of a subdominant
  population of primordial black holes}},
  \href{https://doi.org/10.1103/PhysRevD.98.023536}{\emph{Phys. Rev. D}
  {\bfseries 98} (2018) 023536}
  [\href{https://arxiv.org/abs/1805.09034}{{\ttfamily 1805.09034}}].

\bibitem{LIGOScientific:2019kan}
{\scshape LIGO Scientific, Virgo} collaboration, \emph{{Search for Subsolar
  Mass Ultracompact Binaries in Advanced LIGO\textquoteright{}s Second
  Observing Run}},
  \href{https://doi.org/10.1103/PhysRevLett.123.161102}{\emph{Phys. Rev. Lett.}
  {\bfseries 123} (2019) 161102}
  [\href{https://arxiv.org/abs/1904.08976}{{\ttfamily 1904.08976}}].

\bibitem{Chen:2019irf}
Z.-C.~Chen and Q.-G.~Huang, \emph{{Distinguishing Primordial Black Holes from
  Astrophysical Black Holes by Einstein Telescope and Cosmic Explorer}},
  \href{https://doi.org/10.1088/1475-7516/2020/08/039}{\emph{JCAP} {\bfseries
  08} (2020) 039} [\href{https://arxiv.org/abs/1904.02396}{{\ttfamily
  1904.02396}}].

\bibitem{Boehm:2020jwd}
C.~Boehm, A.~Kobakhidze, C.A.J.~O'hare, Z.S.C.~Picker and M.~Sakellariadou,
  \emph{{Eliminating the LIGO bounds on primordial black hole dark matter}},
  \href{https://doi.org/10.1088/1475-7516/2021/03/078}{\emph{JCAP} {\bfseries
  03} (2021) 078} [\href{https://arxiv.org/abs/2008.10743}{{\ttfamily
  2008.10743}}].

\bibitem{Mena:2019nhm}
O.~Mena, S.~Palomares-Ruiz, P.~Villanueva-Domingo and S.J.~Witte,
  \emph{{Constraining the primordial black hole abundance with 21-cm
  cosmology}}, \href{https://doi.org/10.1103/PhysRevD.100.043540}{\emph{Phys.
  Rev. D} {\bfseries 100} (2019) 043540}
  [\href{https://arxiv.org/abs/1906.07735}{{\ttfamily 1906.07735}}].

\bibitem{Villanueva-Domingo:2021cgh}
P.~Villanueva-Domingo and K.~Ichiki, \emph{{21 cm Forest Constraints on
  Primordial Black Holes}},  \href{https://arxiv.org/abs/2104.10695}{{\ttfamily
  2104.10695}}.

\bibitem{Saito:2008jc}
R.~Saito and J.~Yokoyama, \emph{{Gravitational wave background as a probe of
  the primordial black hole abundance}},
  \href{https://doi.org/10.1103/PhysRevLett.102.161101}{\emph{Phys. Rev. Lett.}
  {\bfseries 102} (2009) 161101}
  [\href{https://arxiv.org/abs/0812.4339}{{\ttfamily 0812.4339}}].

\bibitem{Saito:2009jt}
R.~Saito and J.~Yokoyama, \emph{{Gravitational-Wave Constraints on the
  Abundance of Primordial Black Holes}},
  \href{https://doi.org/10.1143/PTP.126.351}{\emph{Prog. Theor. Phys.}
  {\bfseries 123} (2010) 867}
  [\href{https://arxiv.org/abs/0912.5317}{{\ttfamily 0912.5317}}].

\bibitem{Yuan:2019udt}
C.~Yuan, Z.-C.~Chen and Q.-G.~Huang, \emph{{Probing
  primordial\textendash{}black-hole dark matter with scalar induced
  gravitational waves}},
  \href{https://doi.org/10.1103/PhysRevD.100.081301}{\emph{Phys. Rev. D}
  {\bfseries 100} (2019) 081301}
  [\href{https://arxiv.org/abs/1906.11549}{{\ttfamily 1906.11549}}].

\bibitem{Zhou:2021vcw}
J.-Z.~Zhou, X.~Zhang, Q.-H.~Zhu and Z.~Chang, \emph{{The third order scalar
  induced gravitational waves}},
  \href{https://doi.org/10.1088/1475-7516/2022/05/013}{\emph{JCAP} {\bfseries
  05} (2022) 013} [\href{https://arxiv.org/abs/2106.01641}{{\ttfamily
  2106.01641}}].

\bibitem{Chang:2022nzu}
Z.~Chang, X.~Zhang and J.-Z.~Zhou, \emph{{Primordial black holes and third
  order scalar induced gravitational waves}},
  \href{https://arxiv.org/abs/2209.12404}{{\ttfamily 2209.12404}}.

\bibitem{Ananda:2006af}
K.N.~Ananda, C.~Clarkson and D.~Wands, \emph{{The Cosmological gravitational
  wave background from primordial density perturbations}},
  \href{https://doi.org/10.1103/PhysRevD.75.123518}{\emph{Phys. Rev. D}
  {\bfseries 75} (2007) 123518}
  [\href{https://arxiv.org/abs/gr-qc/0612013}{{\ttfamily gr-qc/0612013}}].

\bibitem{Baumann:2007zm}
D.~Baumann, P.J.~Steinhardt, K.~Takahashi and K.~Ichiki, \emph{{Gravitational
  Wave Spectrum Induced by Primordial Scalar Perturbations}},
  \href{https://doi.org/10.1103/PhysRevD.76.084019}{\emph{Phys. Rev. D}
  {\bfseries 76} (2007) 084019}
  [\href{https://arxiv.org/abs/hep-th/0703290}{{\ttfamily hep-th/0703290}}].

\bibitem{Yuan:2021qgz}
C.~Yuan and Q.-G.~Huang, \emph{{A topic review on probing primordial black hole
  dark matter with scalar induced gravitational waves}},
  \href{https://arxiv.org/abs/2103.04739}{{\ttfamily 2103.04739}}.

\bibitem{Domenech:2021ztg}
G.~Dom\`enech, \emph{{Scalar Induced Gravitational Waves Review}},
  \href{https://doi.org/10.3390/universe7110398}{\emph{Universe} {\bfseries 7}
  (2021) 398} [\href{https://arxiv.org/abs/2109.01398}{{\ttfamily
  2109.01398}}].

\bibitem{Kohri:2018awv}
K.~Kohri and T.~Terada, \emph{{Semianalytic calculation of gravitational wave
  spectrum nonlinearly induced from primordial curvature perturbations}},
  \href{https://doi.org/10.1103/PhysRevD.97.123532}{\emph{Phys. Rev. D}
  {\bfseries 97} (2018) 123532}
  [\href{https://arxiv.org/abs/1804.08577}{{\ttfamily 1804.08577}}].

\bibitem{Hwang:2017oxa}
J.-C.~Hwang, D.~Jeong and H.~Noh, \emph{{Gauge dependence of gravitational
  waves generated from scalar perturbations}},
  \href{https://doi.org/10.3847/1538-4357/aa74be}{\emph{Astrophys. J.}
  {\bfseries 842} (2017) 46}
  [\href{https://arxiv.org/abs/1704.03500}{{\ttfamily 1704.03500}}].

\bibitem{Gong:2019mui}
J.-O.~Gong, \emph{{Analytic Integral Solutions for Induced Gravitational
  Waves}}, \href{https://doi.org/10.3847/1538-4357/ac3a6c}{\emph{Astrophys. J.}
  {\bfseries 925} (2022) 102}
  [\href{https://arxiv.org/abs/1909.12708}{{\ttfamily 1909.12708}}].

\bibitem{Tomikawa:2019tvi}
K.~Tomikawa and T.~Kobayashi, \emph{{Gauge dependence of gravitational waves
  generated at second order from scalar perturbations}},
  \href{https://doi.org/10.1103/PhysRevD.101.083529}{\emph{Phys. Rev. D}
  {\bfseries 101} (2020) 083529}
  [\href{https://arxiv.org/abs/1910.01880}{{\ttfamily 1910.01880}}].

\bibitem{Lu:2020diy}
Y.~Lu, A.~Ali, Y.~Gong, J.~Lin and F.~Zhang, \emph{{Gauge transformation of
  scalar induced gravitational waves}},
  \href{https://doi.org/10.1103/PhysRevD.102.083503(2020)}{\emph{Phys. Rev. D}
  {\bfseries 102} (2020) 083503}
  [\href{https://arxiv.org/abs/2006.03450}{{\ttfamily 2006.03450}}].

\bibitem{Arroja:2009sh}
F.~Arroja, H.~Assadullahi, K.~Koyama and D.~Wands, \emph{{Cosmological matching
  conditions for gravitational waves at second order}},
  \href{https://doi.org/10.1103/PhysRevD.80.123526}{\emph{Phys. Rev. D}
  {\bfseries 80} (2009) 123526}
  [\href{https://arxiv.org/abs/0907.3618}{{\ttfamily 0907.3618}}].

\bibitem{Chang:2020tji}
Z.~Chang, S.~Wang and Q.-H.~Zhu, \emph{{Note on gauge invariance of second
  order cosmological perturbations}},
  \href{https://doi.org/10.1088/1674-1137/ac0c74}{\emph{Chin. Phys. C}
  {\bfseries 45} (2021) 095101}
  [\href{https://arxiv.org/abs/2009.11025}{{\ttfamily 2009.11025}}].

\bibitem{DeLuca:2019ufz}
V.~De~Luca, G.~Franciolini, A.~Kehagias and A.~Riotto, \emph{{On the Gauge
  Invariance of Cosmological Gravitational Waves}},
  \href{https://doi.org/10.1088/1475-7516/2020/03/014}{\emph{JCAP} {\bfseries
  03} (2020) 014} [\href{https://arxiv.org/abs/1911.09689}{{\ttfamily
  1911.09689}}].

\bibitem{Inomata:2019yww}
K.~Inomata and T.~Terada, \emph{{Gauge Independence of Induced Gravitational
  Waves}}, \href{https://doi.org/10.1103/PhysRevD.101.023523}{\emph{Phys. Rev.
  D} {\bfseries 101} (2020) 023523}
  [\href{https://arxiv.org/abs/1912.00785}{{\ttfamily 1912.00785}}].

\bibitem{Yuan:2019fwv}
C.~Yuan, Z.-C.~Chen and Q.-G.~Huang, \emph{{Scalar induced gravitational waves
  in different gauges}},
  \href{https://doi.org/10.1103/PhysRevD.101.063018}{\emph{Phys. Rev. D}
  {\bfseries 101} (2020) 063018}
  [\href{https://arxiv.org/abs/1912.00885}{{\ttfamily 1912.00885}}].

\bibitem{Chang:2020iji}
Z.~Chang, S.~Wang and Q.-H.~Zhu, \emph{{Gauge Invariant Second Order
  Gravitational Waves}},  \href{https://arxiv.org/abs/2009.11994}{{\ttfamily
  2009.11994}}.

\bibitem{Chang:2020mky}
Z.~Chang, S.~Wang and Q.-H.~Zhu, \emph{{On the Gauge Invariance of Scalar
  Induced Gravitational Waves: Gauge Fixings Considered}},
  \href{https://arxiv.org/abs/2010.01487}{{\ttfamily 2010.01487}}.

\bibitem{Domenech:2020xin}
G.~Dom\`enech and M.~Sasaki, \emph{{Approximate gauge independence of the
  induced gravitational wave spectrum}},
  \href{https://doi.org/10.1103/PhysRevD.103.063531}{\emph{Phys. Rev. D}
  {\bfseries 103} (2021) 063531}
  [\href{https://arxiv.org/abs/2012.14016}{{\ttfamily 2012.14016}}].

\bibitem{Cai:2021jbi}
R.-G.~Cai, X.-Y.~Yang and L.~Zhao, \emph{{On the energy of gravitational
  waves}}, \href{https://doi.org/10.1007/s10714-022-02972-x}{\emph{Gen. Rel.
  Grav.} {\bfseries 54} (2022) 89}
  [\href{https://arxiv.org/abs/2109.06864}{{\ttfamily 2109.06864}}].

\bibitem{Cai:2021ndu}
R.-G.~Cai, X.-Y.~Yang and L.~Zhao, \emph{{Energy spectrum of gravitational
  waves}},  \href{https://arxiv.org/abs/2109.06865}{{\ttfamily 2109.06865}}.

\bibitem{Garcia-Bellido:2017aan}
J.~Garcia-Bellido, M.~Peloso and C.~Unal, \emph{{Gravitational Wave signatures
  of inflationary models from Primordial Black Hole Dark Matter}},
  \href{https://doi.org/10.1088/1475-7516/2017/09/013}{\emph{JCAP} {\bfseries
  09} (2017) 013} [\href{https://arxiv.org/abs/1707.02441}{{\ttfamily
  1707.02441}}].

\bibitem{Cai:2018dig}
R.-g.~Cai, S.~Pi and M.~Sasaki, \emph{{Gravitational Waves Induced by
  non-Gaussian Scalar Perturbations}},
  \href{https://doi.org/10.1103/PhysRevLett.122.201101}{\emph{Phys. Rev. Lett.}
  {\bfseries 122} (2019) 201101}
  [\href{https://arxiv.org/abs/1810.11000}{{\ttfamily 1810.11000}}].

\bibitem{Unal:2018yaa}
C.~Unal, \emph{{Imprints of Primordial Non-Gaussianity on Gravitational Wave
  Spectrum}}, \href{https://doi.org/10.1103/PhysRevD.99.041301}{\emph{Phys.
  Rev. D} {\bfseries 99} (2019) 041301}
  [\href{https://arxiv.org/abs/1811.09151}{{\ttfamily 1811.09151}}].

\bibitem{Nakama:2016gzw}
T.~Nakama, J.~Silk and M.~Kamionkowski, \emph{{Stochastic gravitational waves
  associated with the formation of primordial black holes}},
  \href{https://doi.org/10.1103/PhysRevD.95.043511}{\emph{Phys. Rev. D}
  {\bfseries 95} (2017) 043511}
  [\href{https://arxiv.org/abs/1612.06264}{{\ttfamily 1612.06264}}].

\bibitem{Vennin:2015hra}
V.~Vennin and A.A.~Starobinsky, \emph{{Correlation Functions in Stochastic
  Inflation}}, \href{https://doi.org/10.1140/epjc/s10052-015-3643-y}{\emph{Eur.
  Phys. J. C} {\bfseries 75} (2015) 413}
  [\href{https://arxiv.org/abs/1506.04732}{{\ttfamily 1506.04732}}].

\bibitem{Pattison:2017mbe}
C.~Pattison, V.~Vennin, H.~Assadullahi and D.~Wands, \emph{{Quantum diffusion
  during inflation and primordial black holes}},
  \href{https://doi.org/10.1088/1475-7516/2017/10/046}{\emph{JCAP} {\bfseries
  10} (2017) 046} [\href{https://arxiv.org/abs/1707.00537}{{\ttfamily
  1707.00537}}].

\bibitem{Ezquiaga:2019ftu}
J.M.~Ezquiaga, J.~Garc\'\i{}a-Bellido and V.~Vennin, \emph{{The exponential
  tail of inflationary fluctuations: consequences for primordial black holes}},
  \href{https://doi.org/10.1088/1475-7516/2020/03/029}{\emph{JCAP} {\bfseries
  03} (2020) 029} [\href{https://arxiv.org/abs/1912.05399}{{\ttfamily
  1912.05399}}].

\bibitem{Vennin:2020kng}
V.~Vennin, \emph{{Stochastic inflation and primordial black holes}}, Ph.D.
  thesis, U. Paris-Saclay, 6, 2020.
\newblock \href{https://arxiv.org/abs/2009.08715}{{\ttfamily 2009.08715}}.

\bibitem{Figueroa:2020jkf}
D.G.~Figueroa, S.~Raatikainen, S.~Rasanen and E.~Tomberg, \emph{{Non-Gaussian
  Tail of the Curvature Perturbation in Stochastic Ultraslow-Roll Inflation:
  Implications for Primordial Black Hole Production}},
  \href{https://doi.org/10.1103/PhysRevLett.127.101302}{\emph{Phys. Rev. Lett.}
  {\bfseries 127} (2021) 101302}
  [\href{https://arxiv.org/abs/2012.06551}{{\ttfamily 2012.06551}}].

\bibitem{Pattison:2021oen}
C.~Pattison, V.~Vennin, D.~Wands and H.~Assadullahi, \emph{{Ultra-slow-roll
  inflation with quantum diffusion}},
  \href{https://doi.org/10.1088/1475-7516/2021/04/080}{\emph{JCAP} {\bfseries
  04} (2021) 080} [\href{https://arxiv.org/abs/2101.05741}{{\ttfamily
  2101.05741}}].

\bibitem{Figueroa:2021zah}
D.G.~Figueroa, S.~Raatikainen, S.~Rasanen and E.~Tomberg, \emph{{Implications
  of stochastic effects for primordial black hole production in ultra-slow-roll
  inflation}}, \href{https://doi.org/10.1088/1475-7516/2022/05/027}{\emph{JCAP}
  {\bfseries 05} (2022) 027}
  [\href{https://arxiv.org/abs/2111.07437}{{\ttfamily 2111.07437}}].

\bibitem{Animali:2022otk}
C.~Animali and V.~Vennin, \emph{{Primordial black holes from stochastic
  tunnelling}},  \href{https://arxiv.org/abs/2210.03812}{{\ttfamily
  2210.03812}}.

\bibitem{Espinosa:2018eve}
J.R.~Espinosa, D.~Racco and A.~Riotto, \emph{{A Cosmological Signature of the
  SM Higgs Instability: Gravitational Waves}},
  \href{https://doi.org/10.1088/1475-7516/2018/09/012}{\emph{JCAP} {\bfseries
  09} (2018) 012} [\href{https://arxiv.org/abs/1804.07732}{{\ttfamily
  1804.07732}}].

\bibitem{Komatsu_2011}
E.~Komatsu, K.M.~Smith, J.~Dunkley, C.L.~Bennett, B.~Gold, G.~Hinshaw et~al.,
  \emph{Seven-year wilkinson microwave anisotropy probe (wmap*) observations:
  Cosmological interpretation},
  \href{https://doi.org/10.1088/0067-0049/192/2/18}{\emph{The Astrophysical
  Journal Supplement Series} {\bfseries 192} (2011) 18}.

\bibitem{Zhao:2013bba}
W.~Zhao, Y.~Zhang, X.-P.~You and Z.-H.~Zhu, \emph{{Constraints of relic
  gravitational waves by pulsar timing arrays: Forecasts for the FAST and SKA
  projects}}, \href{https://doi.org/10.1103/PhysRevD.87.124012}{\emph{Phys.
  Rev. D} {\bfseries 87} (2013) 124012}
  [\href{https://arxiv.org/abs/1303.6718}{{\ttfamily 1303.6718}}].

\bibitem{LISACosmologyWorkingGroup:2022jok}
{\scshape LISA Cosmology Working Group} collaboration, \emph{{Cosmology with
  the Laser Interferometer Space Antenna}},
  \href{https://arxiv.org/abs/2204.05434}{{\ttfamily 2204.05434}}.

\bibitem{Yagi:2011wg}
K.~Yagi and N.~Seto, \emph{{Detector configuration of DECIGO/BBO and
  identification of cosmological neutron-star binaries}},
  \href{https://doi.org/10.1103/PhysRevD.83.044011}{\emph{Phys. Rev. D}
  {\bfseries 83} (2011) 044011}
  [\href{https://arxiv.org/abs/1101.3940}{{\ttfamily 1101.3940}}].

\bibitem{Bartolo:2018evs}
N.~Bartolo, V.~De~Luca, G.~Franciolini, A.~Lewis, M.~Peloso and A.~Riotto,
  \emph{{Primordial Black Hole Dark Matter: LISA Serendipity}},
  \href{https://doi.org/10.1103/PhysRevLett.122.211301}{\emph{Phys. Rev. Lett.}
  {\bfseries 122} (2019) 211301}
  [\href{https://arxiv.org/abs/1810.12218}{{\ttfamily 1810.12218}}].

\end{thebibliography}\endgroup

\end{document}